\documentclass[11pt]{article}
\usepackage[utf8]{inputenc}
\usepackage{latexsym}
\usepackage[tikz]{bclogo}
\usepackage{tikzsymbols}
\usepackage{amssymb}
\usepackage{amsmath}
\usepackage{amsmath,stackengine}
\usepackage{amsfonts}
\usepackage{cancel}
\usepackage{linearb}
\usepackage[round]{natbib}   
\usepackage{delarray}
\usepackage{mathtools}
\usepackage{harmony}
\usepackage{linearb}
\usepackage{epiolmec}
\usepackage{tikz}
\usepackage{fontawesome}
\usepackage{phaistos}
\usepackage{csvsimple}
\usepackage{blindtext}
\usepackage{lipsum}
\usepackage{adforn}
\usepackage{graphicx}
\usepackage{mathabx}
\usepackage{blindtext}
\usepackage{float}
\usepackage{framed, color}
\usepackage{mdframed}
\usepackage[export]{adjustbox}
\usepackage{mathtools}
\usepackage[normalem]{ulem}
\stackMath
\usepackage{blkarray, bigstrut}
\usepackage{listings}
\usepackage[top=1.5cm,bottom=2cm,left=1cm,right=1cm,asymmetric]{geometry}
\usepackage[export]{adjustbox}
\usepackage{setspace}
\usepackage{nicematrix}
\NiceMatrixOptions{
code-for-first-row = \color{blue} ,
code-for-last-row = \color{blue} ,
code-for-first-col = \color{blue} ,
code-for-last-col = \color{blue}
}
\usepackage{colortbl}
\usepackage{color,soul}
\usepackage{xcolor}
\usepackage{color}
\definecolor{azumi}{rgb}{0.94, 1.0, 1.0}
\definecolor{forest}{rgb}{0.0, 0.5, 0.0}

\usepackage[colorlinks=true,linkcolor=blue, citecolor=blue]{hyperref}
\hypersetup{
    colorlinks=true,
    linkcolor=blue,
    filecolor=blue,      
    urlcolor=blue,
}

\title{}
\author{}
\date{}

\begin{document}

\title{}
\author{Tatiana Daddario \\
Rutgers University \and Richard P. McLean \\
Rutgers University \and Andrew Postlewaite\thanks{%
Postlewaite gratefully acknowledges support from National Science Foundation
grant SES-1851449.} \\
University of Pennsylvania}
\date{}
\title{An Assignment Problem with Interdependent Valuations and Externalities%
\thanks{%
We thank Rakesh Vohra for helpful discussions.}}
\date{March 15, 2023}
\maketitle
\begin{abstract}
In this paper, we take a mechanism design approach to optimal assignment
problems with asymmetrically informed buyers. In addition, the surplus
generated by an assignment of a buyer to a seller may be adversely affected
by externalities generated by other assignments. The problem is complicated
by several factors. Buyers know their own valuations and externality costs
but do not know this same information for other buyers. Buyers also receive
private signals correlated with the state and, consequently, the
implementation problem exhibits interdependent valuations. This precludes a
naive application of the VCG mechanism and to overcome this interdependency
problem, we construct a two-stage mechanism. In the first stage, we exploit
correlation in the firms signals about the state to induce truthful
reporting of observed signals. Given that buyers are honest in stage 1, we
then use a VCG-like mechanism in stage 2 that induces honest reporting of
valuation and externality functions.
\end{abstract}

\newpage

\section{Introduction}

\hspace*{6mm} Suppose a government authority is tasked with allocating
sections of the spectrum to a given set of telecommunications firms. A
central issue in their task concerns externalities: the bandwidth allocated
to a given firm may interfere with other firms' spectrum usage, imposing
negative externalities on those firms. A firm's valuation for different
portions of the spectrum also depends on other factors such as the firm's
demand and operating costs. To complicate matters, these valuations and
externality costs may depend on factors unobserved at the time that economic
transactions take place. For example, there may be industry-wide uncertainty
about regulation, such as the recent \textit{Net Neutrality }debate.%
\footnote{%
Net neutrality is the principle that an internet service provider (ISP) has
to provide access to all sites, content and applications at the same speed,
under the same conditions without blocking or giving preference to any
content.} The value of any part of the spectrum depends substantially on
whether a bill to mandate net neutrality is passed. Consequently, the firms'
valuations for spectrum segments, as well as their externality costs, depend
on a state variable unobservable at the time that economic decisions must be
made.\footnote{%
Similarly, for the problem of allocating airplane gate slots to airlines,
the costs and benefits of assigning a particular type of aircraft departing
from a particular city to a particular gate in another city can depend on
other aircraft-gate pairings thoughout the air traffic system as well as
unobservables like the weather.}

In this paper, we take a mechanism design approach to assignment problems
like those described above. The model will consist of "buyers" and "sellers"
(or "objects") with the goal of assigning buyers to objects so as to
maximize aggregate expected surplus. The surplus enjoyed by a telecom firm
that is allocated a piece of the spectrum is the expected value of that
piece to the firm less the expected interference costs the firm incurs from
the allocation of spectrum pieces to other firms. The authority could
auction off the spectrum pieces but there are difficulties with standard
auction protocols such as first or second price auctions. First, this is not
a standard private values auction problem. A firm's value for a particular
piece of the spectrum depends on what other firms are allocated because of
the externalities.

\hspace*{6mm}The problem is complicated by several factors. We will assume that buyers'
know how their own valuations and the externality costs they incur depend on
the unobservable state but do not know this same information for other
buyers, nor will we posit that a buyer has beliefs about other buyers'
valuations and externality costs. Buyers also receive private signals
correlated with the state (e.g., as a result of their lobbying with respect
to the net neutrality question).\footnote{%
See~\cite{klemperer2004auctions}, Chapter 5.5 for evidence of shared uncertainty about
spectrum values and the sources of the uncertainty.} Consequently, the
problem is an interdependent value problem due to the firms' informative
signals about the state of the world. This precludes a naive application of
the VCG mechanism.

\hspace*{6mm} To overcome the interdependency problem, we construct a
two-stage mechanism. In the first stage, we exploit correlation in the firms
signals about the state. In particular, firms' signals are truthfully
elicited by giving each firm a positive transfer that depends on the
relation of it's announced signal to the other firms' announced signals.\
Given a profile of reported first-stage signals, the mechanism then computes
the associated posterior distribution on the state space and makes this
public.\ In stage 2, firms observe this posted posterior distribution and
then make not necessarily honest reports of their valuations and externality
costs to the mechanism. Once these are reported, the mechanism can identify
a surplus-maximizing assignment. We show that, if buyers honestly report
their observed signals in the first stage, then this second-stage problem is
a private values problem and we are able to use VCG\ transfers in the second
stage to elicit firms' private expected bandwidth valuations and externality
costs. We show that there exist first-stage rewards and second-stage VCG
transfers such that firms honestly report their signals in Stage 1 and their
expected valuations and externality costs in Stage 2 in an equilibrium.
Given the weak informational assumptions of this paper, we propose a
behavior strategy profile and beliefs at each information set that need not
be a perfect Bayesian equilibrium but does satisfy a certain dominant
strategy property along the equilibrium path of play.

\hspace*{6mm} The VCG mechanism used at the second stage is attractive in
that the transfers are non-negative -- any transfers go from the firms to
the authority. It is not surprising that one can structure the VCG mechanism
so that aggregate payments are negative. What is important, though, is that
the outcome of the VCG\ mechanism is individually rational. Again, it is not
surprising that one can construct the VCG mechanism to deliver individually
rational outcomes, but it is \textit{not }generally true that one can
simultaneously satisfy individual rationality and non-positive surplus.

\hspace*{6mm} Offsetting (at least partially)\ the VCG surplus in the second
stage are the first-period rewards that the authority pays to the firms to
elicit the information about the state of the world. If the sum of those
payments is larger than the VCG surplus in the second stage, the authority
would have an aggregate deficit. We show that if there is a deficit, that
deficit goes to zero as the number of buyers and objects goes to infinity.

\section{Related Work}

\hspace*{6mm} Our work is related to matching problems with asymmetric
information.~\cite{fernandez2022centralized} illustrated that even a small
infusion of uncertainty about preferences of other agents into ordinal
matching problems can undermine results achieved under a complete
information framework.~\cite{liu2014stable} elaborated on the notion of
incomplete information stability concept in cardinal matching problems, and
introduced an iterative belief-formation process, used by agents in
allocation blocking decision making. Complementing this work,~\cite%
{roth1989two} points out that some results in a complete information
framework that are related to dominant strategy behavior, can be transferred
with no change to incomplete information framework. In particular, in the
context of marriage problems, every man has a dominant strategy to report
his value truthfully in an M-optimal stable matching mechanism. This
principle holds in both complete and incomplete information frameworks.~\cite%
{demange1985strategy} and~\cite{leonard1983elicitation} draw parallel
conclusions for trading settings under asymmetric information cast as
cardinal matching problems. They show that buyer-optimal stable outcomes can
be supported by VCG transfers.

\hspace*{6mm} This paper is also related to work on auctions with
heterogeneous goods under interdependent information. In the environment of
independent signals and interdependent valuations,~\cite{perry2002efficient}
study multi-unit one-sided auctions, as collections of two-person
single-unit second price auctions. Under a set of assumptions on valuation
functions, including a single-crossing property, Perry and Reny construct a
two-round mechanism, where agents reveal their private signals in the first
round, and, given this aggregated information, estimate their valuations;
while in the second round they engage in corresponding second price
auctions. Our mechanism does not require the assumptions used for Perry and
Reny's results.~\cite{ausubel2006efficient} investigates a dynamic one-sided
auction, for heterogeneous goods, with a single seller, who seeks to
allocate a finite number of goods to multiple potential buyers. He
introduces an adjustment process to calculate agents' allocations and
payments. In this system, the adjustments are driven by aggregate reports of
the opponents, making the truthful reporting a dominant strategy. The
resulting transaction price converges to a competitive Walrasian price.
Unlike Ausubel's framework, which focuses on the trajectory properties of
the model in the time limit, we examine the system's behavior through
replica economies.

\hspace*{6mm} Our model includes allocative externalities. Specifically,
given an assignment, each agent experiences externalities from other
matches. Hence, each agent cares not only about whom he is paired with, but
is also concerned with the other agents' matches. Our definition of
externalities is related to that in~\cite{jehiel1996not}. They consider
allocative externalities which influence the interactions among the agents,
after all contract dealings are concluded. Each player's private information
is his object valuation and the amount of externality he imposes on others.
This formulation creates incentive compatibility issues, which are resolved
via proper threats to the agents. The resulting mechanism includes a
participation stage, and is augmented with an allocation function for cases
when agents decline to participate in the mechanism. We avoid reporting
issues associated with externalities by assuming that each agent knows how
much externality \textit{is experienced by him}, from other agents.
Moreover, we assume that an agent cannot be hurt by the other matches if he
is not assigned the object\footnote{%
This work is tangentially related to assignment-like problems arising in
internet advertising slot allocation or vaccine allocation. Our model is
quite different and is complicated by the presence of interdependent
valuations.}.\newline
\hspace*{6mm} We implement an efficient and asymptotically budget balanced
assignment via a voluntary and incentive compatible mechanism in the
presence of incomplete and interdependent information. Our solution is based
on a two-stage approach developed by~\cite{mclean2017dynamic},~\cite%
{mclean2004informational}. Unlike the aforementioned, in this paper we work
in the environment where the set of feasible outcomes is not fixed and can
vary with the set of players \footnote{%
See discussion in~\cite{narahari2014game}}. The distinguishing feature of this paper is
that the model requires less information on the part of the mechanism
designer and the players. In particular, the agents are not assumed to know
the payoff functions or externalities of other agents. We introduce a new
equilibrium solution concept in dynamic settings to handle this case.

\section{Preliminaries}

\subsection{States, Signals and Payoffs}

If $K$ is a finite set, let $|K|$ denote the cardinality of $K$ and let $%
\Delta (K)$ denote the set of probability measures on $K.$ Let $\Delta
^{\ast }(K)$ denote the subset of $\Delta (K)$ with full support. Throughout
the paper, $||\cdot ||_{2}$ will denote the 2-norm and $||\cdot ||$ will
denote the 1-norm. The real vector spaces on which these norms are defined
will be clear from the context.

We will be concerned with a two sided market consisting of $n$ "buyers" that
are to be matched with $n$ "sellers" or "objects." Let $N=\{1,..,n\}$.
Buyers and sellers will be paired by a mechanism and will then engage in a
transaction. A transaction might involve a transfer of an object, e.g., a
portion of the spectrum to the buyer, or the provision of a service to the
buyer by the seller.

Let $\Theta =\{\theta _{1},..,\theta _{m}\}$ represent a finite set of
states of nature. Neither buyers nor sellers nor the mechanism know the
value of $\theta $ prior to the conclusion of all transactions.

Buyers receive signals correlated with the state and these signals are
private information. Each buyer $i$ receives a signal in a finite set $%
B_{i}. $ Let $B=B_{1}\times \cdots \times B_{n}$ and let $%
b=(b_{1},..,b_{n})\in B$ denote a generic signal profile in $B.$

The surplus generated when buyer $i$ is matched with seller $j$ depends on
the state $\theta \in \Theta ,$ the buyer's valuation for the object, the
cost of assigning the object to the buyer and externalities that adversely
affect the surplus generated by a match of a buyer and a seller.

Buyers care about the seller on the other side of the market with whom they
transact. Consequently, the valuation of buyer $i$ if matched with seller $j$
in state $\theta $ is a nonnegative number $u_{ij}(\theta )$. The
nonnegative cost of assigning seller $j$'s object (or providing seller $j$'s
service) to buyer $i$ also depends on the state $\theta $ and we denote this 
$v_{ij}(\theta ).$

In our model, we allow for the possibility of externalities that negatively
impact the surplus generated by a matching. To incorporate these as a factor
in the implementation problem, let $c_{ij}^{pq}(\theta )$ denote the
externality cost imposed on buyer $i$ if buyer $i$ and seller $j$ are
matched when buyer $p\neq i$ is matched with seller $q\neq j$ and the state
is $\theta .$ The number $c_{ij}^{pq}(\theta )$ is assumed to be
nonnegative. We assume that for all $(i,$ $j)$, $(p,q)$ and $\theta ,$ $%
0\leq u_{ij}(\theta ),v_{ij}(\theta ),c_{ij}^{pq}(\theta )\leq M$ for some $%
M>0.$

Define $c_{ij}(\theta )\in R^{(n-1)(n-1)}$ where $c_{ij}(\theta
)=(c_{ij}^{pq}(\theta ))_{p\neq i},_{q\neq j}.$ We will write $c_{i}(\theta
)=(c_{i1}(\theta ),..,c_{in}(\theta )),$ $c(\theta )=(c_{1}(\theta
),..,c_{n}(\theta ))$ and $c(\pi )=(c_{1}(\pi ),..,c_{n}(\pi ))$ where $%
c_{ij}^{pq}(\pi )=\sum_{\theta \in \Theta }c_{ij}^{pq}(\theta )\pi (\theta
). $ We will write $u_{i}(\theta )=(u_{i1}(\theta ),..,u_{in}(\theta )),$ $%
u(\theta )=(u_{1}(\theta ),..,u_{n}(\theta ))$ and $u_{-i}(\theta
)=(u_{k}(\theta ))_{k\in N\backslash i}.$ Similarly, we write $v_{i}(\theta
)=(v_{i1}(\theta ),..,v_{in}(\theta )),$ $v(\theta )=(v_{1}(\theta
),..,v_{n}(\theta ))$ and $v_{-i}(\theta )=(v_{k}(\theta ))_{k\in
N\backslash i}.$

If $\pi \in \Delta (\Theta ),$ define $u_{i}(\pi )=(u_{i1}(\pi
),..,u_{in}(\pi ))\in \mathbb{R}^{n}$ as $u_{i}(\pi )=\sum_{\theta \in
\Theta }u_{i}(\theta )\pi (\theta )$ and let $u(\pi )=(u_{1}(\pi
),..,u_{n}(\pi )).$ Similarly, define $v_{i}(\pi )=(v_{i1}(\pi
),..,v_{in}(\pi ))$ $\in \mathbb{R}^{n}$ as $v_{i}(\pi )=\sum_{\theta
}v_{i}(\theta )\pi (\theta )$ and let $v(\pi )=(v_{1}(\pi ),..,v_{n}(\pi )).$

\subsection{Stochastic Structure}

Next, let $\Delta ^{\ast }(\Theta \times B)$ denote the set of $P\in \Delta
(\Theta \times B)$ satisfying $P(\theta ,b)>0$ for each $(\theta ,b)\in
\Theta \times B.$ For $P\in \Delta ^{\ast }(\Theta \times B)$ and $b\in B,$
the conditional distribution induced by $P$ on $\Theta $ given $b\in B$ is
denoted $P_{\Theta }(\cdot |b).$ In the interest of notational simplicity,
we will often write $\rho (b)$ for $P_{\Theta }(\cdot |b).$

For each $i$ and $b_{i}\in B_{i},$ the conditional distribution induced by $%
P $ on $B_{-i}$ given $b_{i}\in B$ is denoted $P_{-i}(\cdot |b_{i}).$ That
is, 
\begin{equation*}
P_{-i}(b_{-i}|b_{i})=\sum_{\theta \in \Theta }P(\theta ,b_{-i}|b_{i}).
\end{equation*}

Finally, we assume that $P_{-i}(\cdot |b_{i})\neq P_{-i}(\cdot
|b_{i}^{\prime })$ if $b_{i}\neq b_{i}^{\prime }$.

\subsection{Informational Assumptions}

As stated above, the observed signals of buyers are private information. In
addition, the valuation functions of buyers and their externality functions
are also private information. That is, the valuation-externality function $%
(u_{i}(\theta ),c_{i}(\theta ))_{\theta \in \Theta }$ is known only to buyer 
$i$. On the other hand, the cost functions $(v_{i}(\theta ))_{\theta \in
\Theta }$ as well as the prior $P\in \Delta ^{\ast }(\Theta \times B)$ are
known to all buyers as well as the mechanism. In addition, all buyers, as
well as the mechanism, know the bound $M$ on buyer valuations, buyer
externalities and seller costs.

\subsection{The implementation problem}

An \emph{assignment problem with interdependent values }is a collection $%
(u_{1},..,u_{n},c_{1},..,c_{n},v_{1},..,v_{n},P)=(u,c,v,P)$ where $P\in
\Delta ^{\ast }(\Theta \times B).$ Let $Z$ denote the set of all feasible
matchings of buyers and sellers. That is, $Z$ is the set of all $n\times n$
arrays $z$ such that $z_{ij}\in \{0,1\}$ and 
\begin{equation*}
\text{ }\sum_{i=1}^{n}z_{ij}\leq 1\text{ for all }j\text{ and }%
\sum_{j=1}^{n}z_{ij}\leq 1\text{ for all }i.
\end{equation*}%
The presence of externalities complicates the task of efficiently matching
buyers to sellers in the presence of asymmetric information since buyer $i$
in an $(i,j)$ match incurs externality costs only from those different $%
(p,q) $ matches that actually take place in an optimal assignment.

An \emph{assignment function} is a mapping $F:B\rightarrow Z$ that specifies
an outcome in $Z$ for each profile of announced types. An assignment
function is outcome efficient if for each $b\in B,$ $F(b)$ solves the
quadratic assignment problem%
\begin{equation*}
\text{ }\max_{z\in Z}\sum_{i=1}^{n}\sum_{j=1}^{n}\left[ \sum_{\theta \in
\Theta }\left[ u_{ij}(\theta )-v_{ij}(\theta )-\sum_{p\neq i}\sum_{q\neq
j}c_{ij}^{pq}(\theta )z_{pq}\right] P_{\Theta }(\theta |b)\right] z_{ij}.
\end{equation*}

Given a matching problem, our goal in this paper is to implement an outcome
efficient assignment function via an individually rational, incentive
compatible mechanism. Furthermore, we wish to design the mechanism so as to
require as little information as possible on the part of buyers and the
mechanism designer. To accomplish this, we present a two-stage mechanism
that borrows its structure from~\cite{mclean2017dynamic}. To simplify the
presentation that follows, we introduce one more piece of notation. For each 
$i$, let $w_{i}=(w_{i1},..,w_{in})\in \mathbb{R}^{n}\ $(interpreted as a
valuation vector)$,d_{i}=(d_{i1},..,d_{in})$ where each $d_{ij}\in \mathbb{R}%
_{+}^{(n-1)(n-1)}$ (interpreted as an externality vector) and $\kappa
_{i}=(\kappa _{i1},..,\kappa _{in})\in \mathbb{R}^{n}$(interpreted as a
seller's cost vector). For each $i,j\in N$ and $z\in Z,$ define buyer $%
i^{\prime }s$ payoff for a given assignment $z\in Z$ as 
\begin{equation*}
g_{i}(z;w_{i},d_{i},\kappa _{i})=\sum_{j=1}^{n}\left[ w_{ij}-\sum_{p\neq
i}\sum_{q\neq j}d_{ij}^{pq}z_{pq}-\kappa _{ij}\right] z_{ij}.
\end{equation*}%
Note that 
\begin{equation*}
|g_{i}(z;w_{i},d_{i},\kappa _{i})|\leq (2+n)M.
\end{equation*}%
In this notation, an assignment function is outcome efficient if for each $%
b\in B,$ $F(b)$ solves the quadratic assignment problem%
\begin{equation*}
\text{ }\max_{z\in Z}\sum_{i=1}^{n}\left[ \sum_{\theta \in \Theta
}g_{i}(z;u_{i}(\theta ),c_{i}(\theta ),v_{i}(\theta ))P_{\Theta }(\theta |b)%
\right]
\end{equation*}

\section{The two-stage Implementation Game}

\subsection{Preliminaries}

We begin by considering a simpler implementation problem that will provide
the basic structure of the second stage of the more complex two-stage model
to follow. The mechanism seeks to match buyers and sellers so as to maximize
total surplus generated by their actual valuations. We will assume that the
seller cost profile $\kappa =(\kappa _{1},..,\kappa _{n})$\ is known to all
buyers and the mechanism but that $w_{i}$\ and $d_{i}$ are known only to
buyer $i$. In this simple problem, each buyer makes a (not necessarily
honest) report of his valuation-externality vector to the mechanism. If $%
\kappa =(\kappa _{1},..,\kappa _{n})$ is a fixed profile of seller costs and
if $(w,d)=(w_{1},..,w_{n},d_{1},..,d_{n})$\ is the profile of buyers'
reports, the mechanism then computes payoffs and classical VCG transfers for
the buyers by solving the quadratic assignment problem that maximizes total
surplus. We show that these transfers yield a mechanism that is dominant
strategy incentive compatible, individually rational and incurs no budget
deficit. Suppressing the dependence of the optimal solution on $\kappa ,$
define for each $(w,d)$ the outcome 
\begin{equation*}
\hat{\varphi}(w,d)\in \arg \text{ }\max_{z\in
Z}\sum_{i}g_{i}(z;w_{i},d_{i},\kappa _{i}).
\end{equation*}

The resulting payoff to buyer $i$ in the optimal solution measured as "net
benefit" is then 
\begin{equation*}
g_{i}(\hat{\varphi}(w,d);w_{i},d_{i},\kappa _{i})=\sum_{j=1}^{n}\left[
w_{ij}-\sum_{p\neq i}\sum_{q\neq j}d_{ij}^{pq}\hat{\varphi}_{pq}(w,d)-\kappa
_{ij}\right] \hat{\varphi}_{ij}(w,d)
\end{equation*}%
while the total payoff to buyers different from $i$ is 
\begin{equation*}
\text{ }\sum_{k:k\neq i}g_{k}(\hat{\varphi}(w,d);w_{k},d_{k},\kappa
_{k})=\sum_{k:k\neq i}\sum_{j=1}^{n}\left[ w_{kj}-\sum_{p\neq k}\sum_{q\neq
j}d_{kj}^{pq}\hat{\varphi}_{pq}(w,d)-\kappa _{kj}\right] \hat{\varphi}%
_{kj}(w,d).
\end{equation*}%
Next, let $Z_{-i}=\{z\in Z|z_{ij}=0$ for all $j\}$ and note that $%
Z_{-i}\subseteq Z$ . Define transfers as follows:%
\begin{equation*}
x_{i}(w,d)=\text{ }\sum_{k:k\neq i}g_{k}(\hat{\varphi}(w,d);w_{k},d_{k},%
\kappa _{k})-\max_{z\in Z_{-i}}\left[ \text{ }\sum_{k:k\neq
i}g_{k}(z;w_{k},d_{k},\kappa _{k})\right] .
\end{equation*}%
Since these transfers define a standard Groves mechanism, dominant strategy
incentive compatibility is immediately obtained. That is, for any buyer $i$
and any valuation-externality vectors $(w_{i},d_{i})$ and $(w_{i}^{\prime
},d_{i}^{\prime })$ and any profile of valuation-externality vectors $%
(w_{-i},d_{-i})$ of other buyers we have%
\begin{equation*}
g_{i}(\hat{\varphi}(w_{-i},d_{-i},w_{i},d_{i});w_{i},d_{i},\kappa
_{i})+x_{i}(w_{-i},d_{-i},w_{i},d_{i})\geq g_{i}(\hat{\varphi}%
(w_{-i},d_{-i},w_{i}^{\prime },d_{i}^{\prime });w_{i},d_{i},\kappa
_{i})+x_{i}(w_{-i},d_{-i},w_{i}^{\prime },d_{i}^{\prime }).
\end{equation*}

\bigskip

To verify individual rationality, suppose that 
\begin{equation*}
z^{\ast }\in \arg \max_{z\in Z_{-i}}\left[ \text{ }\sum_{k:k\neq
i}g_{k}(z;w_{k},d_{k},\kappa _{k})\right] .
\end{equation*}%
Since $z^{\ast }\in Z$ and $g_{i}(z^{\ast };w_{i},d_{i},\kappa _{i})=0,$ it
follows that 
\begin{eqnarray*}
g_{i}(\hat{\varphi}(w,d);w_{i},d_{i},\kappa _{i})+x_{i}(w,d)
&=&\sum_{k=1}^{n}g_{k}(\hat{\varphi}(w,d);w_{k},d_{k},\kappa _{k})-\left[ 
\text{ }\sum_{k:k\neq i}g_{k}(z^{\ast };w_{k},d_{k},\kappa _{k})\right] \\
&=&\sum_{k=1}^{n}g_{k}(\hat{\varphi}(w,d);w_{k},d_{k},\kappa _{k})-\left[ 
\text{ }\sum_{k=1}^{n}g_{k}(z^{\ast };w_{k},d_{k},\kappa _{k})\right] \\
&\geq &0.
\end{eqnarray*}

To verify that transfers $x_{i}(w,d)$ are nonpositive, let $z_{kj}^{\prime }=%
\hat{\varphi}_{kj}(w,d)$ for all $j$ and all $k\neq i$ and $z_{ij}^{\prime
}=0$ for all j. Then $z^{\prime }\in Z_{-i}.$ Since 
\begin{equation*}
\sum_{q\neq j}d_{kj}^{iq}\varphi _{iq}(w,d)\geq 0
\end{equation*}%
we conclude that 
\begin{eqnarray*}
&&\sum_{k:k\neq i}g_{k}(\hat{\varphi}(w,d);w_{k},d_{k},\kappa _{k}) \\
&=&\sum_{k\neq i}\sum_{j}\left[ w_{kj}-\kappa _{kj}-\sum_{p\neq
k,i}\sum_{q\neq j}d_{kj}^{pq}\hat{\varphi}_{pq}(w,d)-\sum_{q\neq
j}d_{kj}^{iq}\hat{\varphi}_{iq}(w,d)\right] \hat{\varphi}_{kj}(w,d) \\
&\leq &\sum_{k\neq i}\sum_{j}\left[ w_{kj}-\kappa _{kj}-\sum_{p\neq
k}\sum_{q\neq j}d_{kj}^{pq}z_{pq}^{\prime }\right] z_{kj}^{\prime } \\
&\leq &\max_{z\in Z_{-i}}\left[ \text{ }\sum_{k:k\neq
i}g_{k}(z;w_{k},d_{k},\kappa _{k})\right]
\end{eqnarray*}%
and transfers are nonpositive.

\subsection{A two-stage game}

We wish to formulate our implementation problem with interdependent
valuations as a two-stage problem in which honest reporting of the agents'
signals in stage one resolves the \textquotedblleft
interdependency\textquotedblright\ problem so that the second-stage problem
is a simple implementation problem with private values of the type presented
in Section 2.1\ above, to which the VCG mechanism can be immediately
applied. We now define an extensive form game that formalizes the two-stage
game that lies behind this idea. Let $\xi =(\xi _{i})_{i\in N}$ be an $n$%
-tuple of functions $\xi _{i}:B\rightarrow \mathbb{R}_{+}$ each of which
assigns to each $b\in B$ a nonnegative number $\xi _{i}(b)$\ interpreted as
a \textquotedblleft reward\textquotedblright\ to buyer $i$. These rewards
are designed to induce buyers to honestly report their signals in stage 1.
We now describe the extensive form of a two-stage game $\Gamma (u,c,v,P,\xi
).$

\bigskip

\textbf{Stage 1}: Buyer $i$ learns his signal $b_{i}$ and let $b\in B$
denote the resulting profile of observed signals. Buyer $i$ then makes a
(not necessarily honest) report of $r_{i}$ to the mechanism. If $%
r=(r_{1},..,r_{n})$ is the profile of stage 1 reports, then buyer $i$
receives a nonnegative payment $\xi _{i}(r),$ and the game moves to stage 2.

\bigskip

\textbf{Stage 2}: If $r$ is the reported signal profile in stage 1, the
mechanism publicly posts the conditional distribution $P_{\Theta }(\cdot
|r)=\rho (r)$ but not the reported profile $r.$ Upon observing $\rho (r),$
each buyer $i$ reports a valuation-externality vector $(w_{i}(\theta
),d_{i}(\theta ))_{\theta \in \Theta }$ to the mechanism. Given the reported
profile $(w(\theta ),d(\theta ))_{\theta \in \Theta }=(w_{i}(\theta
),d_{i}(\theta ))_{\theta \in \Theta ,i\in N}$ and the posted $\rho (r),$
the mechanism then computes ($w_{k}(\rho (r)),d_{k}(\rho (r)),v_{k}(\rho
(r))=\sum_{\theta }(w_{k}(\theta ),d_{k}(\theta ),v_{k}(\theta ))P(\theta
|r) $ for each buyer $k.$ Next, the mechanism computes the social outcome%
\begin{equation*}
\hat{\varphi}(w(\rho (r)),d(\rho (r)))\in \arg \text{ }\max_{z\in
Z}\sum_{k=1}^{n}g_{k}(z;w_{k}(\rho (r)),d_{k}(\rho (r)),v_{k}(\rho (r)))
\end{equation*}%
and VCG transfers $x_{i}(w(\rho (r)),d(\rho (r)))$ for each buyer $i$ where%
\begin{eqnarray*}
&&x_{i}(w(\rho (r)),d(\rho (r)))=\sum_{k:k\neq i}g_{k}(\hat{\varphi}(w(\rho
(r)),d(\rho (r)));w_{k}(\rho (r)),d_{k}(\rho (r)),v_{k}(\rho (r))) \\
&&-\max_{z\in Z_{-i}}\left[ \text{ }\sum_{k:k\neq i}g_{k}(z;w_{k}(\rho
(r)),d_{k}(\rho (r)),v_{k}(\rho (r)))\right] .
\end{eqnarray*}%
If $\hat{\varphi}_{ij}(w(\rho (r)),d(\rho (r)))=1,$ then buyer $i$ receives
transfer $x_{i}(w(\rho (r)),d(\rho (r))),$ $i$ pays seller $j$ the amount $%
v_{ij}(\rho (r))$ and $j$'s object is given (or $j$'s service is provided)
to buyer $i$.

Since $\rho (b)=P_{\Theta }(\cdot |b)$ is the conditional distribution on
states given the realized signal profile $b$, the resulting expected payoff
to buyer $i$ at the end of stage 2 (but before the state $\theta $ is known)
is%
\begin{equation*}
g_{i}(\hat{\varphi}(w(\rho (r)),d(\rho (r)));u_{i}(\rho (b)),c_{i}(\rho
(b)),v_{i}(\rho (r)))+x_{i}(w(\rho (r)),d(\rho (r))+\xi _{i}(r).
\end{equation*}

We wish to design the rewards $\xi _{i}$ so as to accomplish two goals. In
stage 1, we want to induce agents to report honestly so that the reported
stage 1 profile is exactly $b$ when the realized signal profile is $b$.
Then, upon observing the posted posterior distribution $P_{\Theta }(\cdot
|b)=\rho (b)$ in stage 2, we want each buyer $i$ to report the true
valuation-externality vector $(u_{i}(\theta ),c_{i}(\theta )_{\theta \in
\Theta }.$ If these twin goals are accomplished in an equilibrium, then the
social outcome is%
\begin{equation*}
\hat{\varphi}(u(\rho (b)),c(\rho (b)))\in \arg \max_{z\in
Z}\sum_{i}g_{i}(z;u_{i}(\rho (b)),c_{i}(\rho (b)),v_{i}(\rho (b))).
\end{equation*}%
In particular,%
\begin{equation*}
\hat{\varphi}(u(\rho (b)),c(\rho (b)))\in \arg \text{ }\max_{z\in Z}\sum_{i}%
\left[ \sum_{\theta \in \Theta }g_{i}(z;u_{i}(\theta ),c_{i}(\theta
),v_{i}(\theta ))P_{\Theta }(\theta |b)\right]
\end{equation*}%
so that $b\mapsto $ $\hat{\varphi}(u(\rho (b)),c(\rho (b)))$ is an outcome
efficient assignment.

\bigskip

The resulting nonnegative expected payoff to buyer $i$ at the end of stage 2
(but before the state $\theta $ is known) is 
\begin{equation*}
g_{i}(\hat{\varphi}(u(\rho (b)),c(\rho (b)));u_{i}(\rho (b)),c_{i}(\rho
(b)),v_{i}(\rho (b)))+x_{i}(u(\rho (b)),c(\rho (b)))+\xi _{i}(b)
\end{equation*}%
and the net payment to the mechanism is%
\begin{equation*}
\sum_{i}x_{i}(u(\rho (b)),c(\rho (b)))+\sum_{i}\xi _{i}(b).
\end{equation*}%
While the sum of VCG transfers is nonpositive, this net payment could be
positive since the first-stage rewards are nonnegative. Consequently, it is
important to identify situations in which the sum of the first-period
rewards is small and this issue is considered in Section 5 below in the case
of large n.

\subsection{Strategies and Equilibria in the two-stage Game}

For each buyer $i$ and each $b_{i}\in B_{i},$ let 
\begin{equation*}
D(b_{i}):=\{P_{\Theta }(\cdot |b_{-i},b_{i}):b_{-i}\in B_{-i}\}\text{ }
\end{equation*}%
and define a partition $\Pi _{i}(b_{i})=D(b_{i})/\sim $ where $P_{\Theta
}(\cdot |b_{-i}^{\prime },b_{i})\sim P_{\Theta }(\cdot |b_{-i}^{\prime
\prime },b_{i})$ if and only if $P_{\Theta }(\cdot |b_{-i}^{\prime
},b_{i})=P_{\Theta }(\cdot |b_{-i}^{\prime \prime },b_{i}).$

Given the specification of the extensive form, the second-stage information
sets of buyer $i$ are indexed by triples $(r_{i},\pi ,b_{i})$ where $%
b_{i}\in B_{i}$ is the privately observed signal of buyer $i$ in stage 1, $%
r_{i}\in B_{i}$ is the reported signal of buyer $i$ in stage 1, and $\pi \in
\Pi _{i}(r_{i})\}$ is the posted posterior distribution. Consequently, a
behavior strategy for buyer $i$ in this game is a pair $(\alpha _{i},\beta
_{i})$ where $b_{i}\in B_{i}\mapsto \alpha _{i}(b_{i})\in B_{i}$ specifies a
first-stage report as a function of $i$'s observed signal $b_{i}.$ At each
second-stage information set, $\beta _{i}$ specifies for each $\theta $ and $%
j$ a valuation $w_{ij}(\theta )$ and an externality vector $d_{ij}(\theta
)\in \mathbb{R}^{(n-1)(n-1)}.$ More formally, $\beta _{i}$ is a function 
\begin{equation*}
(r_{i},\pi ,b_{i})\in B_{i}\times \Pi _{i}(r_{i})\times B_{i}\mapsto \beta
_{i}(r_{i},\pi ,b_{i})\in (\mathbb{R}^{n}\times \mathbb{R}^{n(n-1)(n-1)})^{m}
\end{equation*}%
that specifies a second-stage report $([\beta _{i}(r_{i},\pi ,b_{i})](\theta
))_{\theta \in \Theta }$ as a function of $i$'s private signal $b_{i}$ $\in
B_{i},$ $i$'s first-stage report $r_{i}$ and the posted distribution $\pi
\in \Pi _{i}(r_{i}).$

\bigskip

We are interested in an equilibrium assessment for the two-stage
implementation game consisting of a strategy profile $(\alpha _{i},\beta
_{i})_{i\in N}$ and a system of second-stage beliefs for each buyer $i$n
which buyers truthfully report their private information at each stage.

\bigskip

\textbf{Definition 1}: A strategy $(\alpha _{i},\beta _{i})$ for buyer $i$
is \emph{truthful} for $i$ in $\Gamma (u,c,v,P,\xi )$ if $\alpha
_{i}(b_{i})=b_{i}$ for all $b_{i}\in B_{i}$ and 
\begin{equation*}
\beta _{i}(b_{i},\pi ,b_{i})=(u_{i}(\theta ),c_{i}(\theta ))_{\theta \in
\Theta }
\end{equation*}%
for all $b_{i}\in B_{i}$ and all $\pi \in \Pi _{i}(b_{i}).$

\bigskip

A strategy profile $(\alpha _{i},\beta _{i})_{i\in N}$ is \emph{truthful} in 
$\Gamma (u,c,v,P,\xi )$ if $(\alpha _{i},\beta _{i})$ is truthful for each
buyer $i$. In a truthful strategy, each buyer $i$ honestly reports the
observed signal in stage 1. Then, in stage 2, buyer $i$ honestly reports his
true valuation-externality function at any second-stage information set
corresponding to a reported signal that matches i's observed signal in stage
1.

Formally, a \textbf{system of beliefs} for buyer $i$ is a collection of
probability measures in $\Delta (\Theta \times B_{-i})$ indexed by $%
(r_{i},\pi ,b_{i})$ with $\pi \in \Pi _{i}(r_{i})$ where $\mu _{i}(\cdot
|r_{i},\pi ,b_{i})\in \Delta (\Theta \times B_{-i})$ has the following
interpretation: when player $i$ observes signal $b_{i}$ reports $r_{i}$ in
Stage 1 and observes the posted distribution $\pi \in \Pi _{i}(r_{i})$ in
stage 2, then buyer $i$ assigns probability mass $\mu _{i}(\theta
,b_{-i}|r_{i},\pi ,b_{i})$ to the event that other buyers have observed
signals $b_{-i}$ and the state of nature is $\theta .$ As usual, an \emph{%
assessment} is a collection $\{(\alpha _{i},\beta _{i})_{i\in N}$ $,(\mu
_{i})_{i\in N}\}$ consisting of a behavior strategy and a system of beliefs
for each buyer $i$.\newline

\bigskip

\textbf{Definition 2}: An assessment $\{(\alpha _{i}^{\ast },\beta
_{i}^{\ast })_{i\in N}$ $,(\mu _{i}^{\ast })_{i\in N}\}$ is a \emph{%
sequential dominant strategy assessment} in $\Gamma (u,,c,v,P,\xi )$ if $%
\{(\alpha _{i}^{\ast },\beta _{i}^{\ast })_{i\in N}$ $,(\mu _{i}^{\ast
})_{i\in N}\}$ is a Nash equilibrium assessment satisfying (i) the profile $%
((\alpha _{i}^{\ast },\beta _{i}^{\ast })_{i\in N}$ is truthful,\ (ii) for
any profile $(\alpha _{j},\beta _{j})_{j\in N\backslash i}$ of behavior
strategies of other players, $(\alpha _{i}^{\ast },\beta _{i}^{\ast })$ is
sequentially rational for $i$ at each first-stage information set $b_{i}\in
B_{i}$ and (iii) for each $b_{i}\in B_{i}$ and each $\pi \in \Pi _{i}(b_{i})$
and for any profile $(\beta _{j})_{j\in N\backslash i}$ of second-stage
strategies of other players, $(\alpha _{i}^{\ast },\beta _{i}^{\ast })$ is
sequentially rational at each second-stage information set given $(\alpha
_{j}^{\ast },\beta _{j})_{j\in N\backslash i}.$

\bigskip

Informally, $\{(\alpha _{i}^{\ast },\beta _{i}^{\ast })_{i\in N}$ $,(\mu
_{i}^{\ast })_{i\in N}\}$ is a sequential dominant strategy assessment if
for each $i$, ($\alpha _{i}^{\ast },\beta _{i}^{\ast })$ is a dominant
strategy "along the equilibrium path of play." Condition (ii) states that, $%
(\alpha _{i}^{\ast },\beta _{i}^{\ast })$ is a best response against any $%
(\alpha _{j},\beta _{j})_{j\in N\backslash i}$ at each first-stage
information set and condition (iii) states that $(\alpha _{i}^{\ast },\beta
_{i}^{\ast })$ is a best response against any $(\alpha _{j}^{\ast },\beta
_{j})_{j\in N\backslash i}$ given that all players use $(\alpha _{j}^{\ast
})_{j\in N}$ in stage 1 and beliefs are derived from Bayes rule.

\subsection{The Main Result}

\textbf{Theorem 1}: Let $(u,c,v,P)$ be an assignment problem with
interdependent valuations. For each i, define a behavior strategy $(\alpha
_{i}^{\ast },\beta _{i}^{\ast })$ for $i$ where $\alpha _{i}^{\ast
}(b_{i})=b_{i}$ for each $b_{i}\in B_{i}$ and $\beta _{i}^{\ast }(r_{i},\pi
,b_{i})=(u_{i}(\theta ),c_{i}(\theta ))_{\theta \in \Theta }$ for each $%
((r_{i},\pi ,b_{i})$ such that $r_{i}\in B_{i},\pi \in \Pi _{i}(r_{i})$ and $%
b_{i}\in B_{i}.$\ Then there exists a reward system $\xi =(\xi _{i})_{i\in
N} $ and second-stage beliefs $(\mu _{i}^{\ast })_{i\in N}$ for each $i$such
that $(\alpha _{i}^{\ast },\beta _{i}^{\ast },\mu _{i}^{\ast })_{i\in N}$ is
a sequential dominant strategy assessment in the two-stage game $\Gamma
(u,c,v,P,\xi ).$

\bigskip

To prove this result, we proceed in several steps which we outline here. Let 
$(\alpha _{i}^{\ast },\beta _{i}^{\ast })_{i\in N}$ be defined as in the
statement of Theorem 1. Clearly, $(\alpha _{i}^{\ast },\beta _{i}^{\ast
})_{i\in N}$ is a truthful profile.

\bigskip

\emph{Step 1}: A equilibrium assessment requires that second-stage beliefs
be specified for each buyer $i$ at each of buyer $i$'s second-stage
information sets. Furthermore, given $(\alpha _{i}^{\ast },\beta _{i}^{\ast
})_{i\in N},$ these should be "consistent with Bayes rule whenever
possible." Suppose that buyer $i$ receives signal $b_{i},$ the other players
receive signal profile $b_{-i}\in B_{-i},$ and buyer $i$ reports $r_{i}$ in
stage 1. Then $\alpha _{k}^{\ast }(b_{k})=b_{k}$ for each $k\neq i$.
Therefore, buyer $i$ with signal $b_{i}$ who has submitted report $r_{i}$ in
stage 1 and who observes $\pi \in \Pi (r_{i})$ at stage 2 will assign
positive probability 
\begin{equation*}
\sum_{\hat{b}_{-i}:\rho (\hat{b}_{-i},r_{i})=\pi }P_{-i}(\hat{b}_{-i}|b_{i})
\end{equation*}%
to the event 
\begin{equation*}
\{\hat{b}_{-i}\in B_{-i}:\rho (\hat{b}_{-i},r_{i})=\pi \}.
\end{equation*}%
Therefore, $i$'s updated beliefs regarding $(\theta ,b_{-i})$ consistent
with $(\alpha ^{\ast },\beta ^{\ast })$ are given by 
\begin{eqnarray*}
\mu _{i}^{\ast }(\theta ,b_{-i}|r_{i},\pi ,b_{i}) &=&\frac{P_{\Theta
}(\theta |b_{-i},b_{i})P_{-i}(b_{-i}|b_{i})}{\sum_{\hat{b}_{-i}:\rho (\hat{b}%
_{-i},r_{i})=\pi }P_{-i}(\hat{b}_{-i}|b_{i})}\text{ if }\rho
(b_{-i},r_{i})=\pi \\
&=&0\text{ otherwise.}
\end{eqnarray*}

\emph{Step 2}: To verify part (iii) of the definition of sequential dominant
strategy assessment, we must show that, if all buyers are truthful in stage
1, then $\beta _{i}^{\ast }$ is a best response against any collection $%
(\beta _{j})_{j\neq i}$ of second-stage strategies of i's opponents. For
each $b_{-i}\in B_{-i}$ and each $k\neq i,$ let 
\begin{equation*}
\sum_{\theta }[\beta _{k}(b_{k},\pi ,b_{k})](\theta )\pi (\theta )=\gamma
_{k}(\pi ,b_{k})
\end{equation*}%
and 
\begin{equation*}
(\gamma _{j}(\pi ,b_{j}))_{j\neq i}=\gamma _{-i}(\pi ,b_{-i}).
\end{equation*}

To verify part (iii), we show that for each $b_{i}\in B_{i}$ and $\pi \in
\Pi (b_{i})$ and each $(w_{i}(\theta ),d_{i}(\theta ))_{\theta \in \Theta },$
we have 
\begin{eqnarray*}
&&\sum_{b_{-i}\in B_{-i}}\sum_{\theta \in \Theta }\left[ g_{i}(\hat{\varphi}%
(\gamma _{-i}(\pi ,b_{-i}),u_{i}(\pi ),c_{i}(\pi ));u_{i}(\theta
),c_{i}(\theta ),v_{i}(\pi )\right] +\hat{x}_{i}(\gamma _{-i}(\pi
,b_{-i}),u_{i}(\pi ),c_{i}(\pi )))\mu _{i}(\theta ,b_{-i}|b_{i},\pi ,b_{i})
\\
&\geq &\sum_{b_{-i}\in B_{-i}}\sum_{\theta \in \Theta }\left[ g_{i}(\hat{%
\varphi}(\gamma _{-i}(\pi ,b_{-i}),w_{i}(\pi ),d_{i}(\pi ));u_{i}(\theta
),c_{i}(\theta ),v_{i}(\pi )\right] +\hat{x}_{i}(\gamma _{-i}(\pi
,b_{-i}),w_{i}(\pi ),d_{i}(\pi )))\mu _{i}(\theta ,b_{-i}|b_{i},\pi ,b_{i}).
\end{eqnarray*}

\bigskip

\emph{Step 3}: To verify part (iii) of the definition of sequential dominant
strategy assessment, we must show that ($(\alpha ^{\ast },\beta ^{\ast
}),\mu ^{\ast })$ is sequentially rational for buyer $i$ at his first-stage
information sets given any behavior strategy profile $(\alpha _{k},\beta
_{k})_{k\neq i}$ of the other players. In particular, we must show that a
coordinated deviation by buyer $i$ in which buyer $i$ lies in stage 1 and
reports optimally in the second stage given the first-stage lie, is not
profitable when the sellers different from $i$ use any behavior strategy
profile $((\alpha _{k},\beta _{k})_{k\neq i}.$ In this step, the first-stage
rewards play a crucial role.

To see this, choose $b=(b_{-i},b_{i}),$ $r_{i}\in B_{i}$ and let $\pi =\rho
(b)$ and $\pi ^{\prime }=\rho (b_{-i},r).$ For each $j\neq i,$ let 
\begin{equation*}
\sum_{\theta \in \Theta }[\beta _{j}(\alpha _{j}(b_{j}),\pi ,b_{j})](\theta
)\pi (\theta )=\gamma _{j}(\pi ,b_{j})
\end{equation*}%
and 
\begin{equation*}
\sum_{\theta \in \Theta }[\beta _{j}(\alpha _{j}(b_{j}),\pi ^{\prime
},b_{j})](\theta )\pi ^{\prime }(\theta )=\gamma _{j}(\pi ^{\prime },b_{j}).
\end{equation*}%
Let $(\gamma _{j}(\pi ,b_{j}))_{j\neq i}=\gamma _{-i}(\pi ,b_{-i})$ and $%
(\gamma _{j}(\pi ^{\prime },b_{j}))_{j\neq i}=\gamma _{-i}(\pi ^{\prime
},b_{-i}).$ We first show that, for all $(w_{i}(\theta ),d_{i}(\theta
))_{\theta \in \Theta },$ we have%
\begin{eqnarray*}
&&g_{i}(\hat{\varphi}(\gamma _{-i}(\pi ,b_{-i}),u_{i}(\pi ),c_{i}(\pi
));u_{i}(\pi ),c_{i}(\pi ),v_{i}(\pi ))+\hat{x}_{i}(\gamma _{-i}(\pi
,b_{-i}),u_{i}(\pi ),c_{i}(\pi )) \\
&&-\max_{(w_{i}(\theta ),d_{i}(\theta ))_{\theta \in \Theta }}[g_{i}(\hat{%
\varphi}(\gamma _{-i}(\pi ^{\prime },b_{-i}),w_{i}(\pi ^{\prime }),d_{i}(\pi
^{\prime }));u_{i}(\pi ),c_{i}(\pi ),v_{i}(\pi ^{\prime }))+\hat{x}%
_{i}(\gamma _{-i}(\pi ^{\prime },b_{-i}),w_{i}(\pi ^{\prime }),d_{i}(\pi
^{\prime }))] \\
&\geq &-(2+n)(2n+1)M||\pi -\pi ^{\prime }|| \\
&=&-(2+n)(2n+1)M||\rho (b)-\rho (b_{-i},r_{i})||.
\end{eqnarray*}%
Consequently, a first-stage deviation will be unprofitable if 
\begin{equation*}
-(2+n)(2n+1)M\sum_{b_{-i}}||\rho (b)-\rho
(b_{-i},r_{i})||P_{-i}(b_{-i}|b_{i})+\sum_{b_{-i}}{\LARGE [}\xi _{i}(b)-\xi
_{i}(b_{-i},r_{i}){\LARGE ]}P_{-i}(b_{-i}|b_{i})>0.
\end{equation*}

To complete the proof , we construct a reward system $(\xi _{i})_{i\in N}$
defined by spherical scoring rules: for each $b$, and $i$, 
\begin{equation*}
\xi _{i}(b)=\delta \frac{P_{-i}(b_{-i}|b_{i})}{||P_{-i}(\cdot |b_{i})||_{2}}.
\end{equation*}%
Since 
\begin{equation*}
\sum_{b_{-i}\in B_{-i}}{\LARGE [}\xi _{i}(b)-\xi _{i}(b_{-i},r_{i}){\LARGE ]}%
P_{-i}(b_{-i}|b_{i}))>0
\end{equation*}%
we can choose $\delta $ so as to ensure that all agents report their
first-stage signals honestly.

\subsection{Remarks}

Unlike the previous paper,~\cite{mclean2017dynamic}, we do not claim that
the sequential dominant strategy assessment $(\alpha ^{\ast },\beta ^{\ast
},\mu ^{\ast })$ defined in Theorem 1 is a perfect Bayesian equilibrium
assessment. Note that for $r_{i}\neq b_{i},$ the equilibrium path of play
does not pass through the information sets indexed by $(r_{i},\pi ,b_{i})$
with $\pi \in \Pi _{i}(r_{i}).$ There is no difficulty in identifying
beliefs at these information sets compatible with Bayes rule and we have
computed $\mu _{i}^{\ast }(\theta ,b_{-i}|r_{i},\pi ,b_{i})$ above. There is
also no difficulty in defining second-stage strategies at these information
sets and we define $\beta _{i}^{\ast }(r_{i},\pi ,b_{i})=(u_{i}(\theta
),c_{i}(\theta ))_{\theta \in \Theta }$ as above. The difficulty arises when 
$i$ wishes to compute his best response at one of these unreached
information sets. If $i$ knew $(u_{j}(\theta ),c_{j}(\theta ))_{j\neq i}$
(as in~\cite{mclean2017dynamic}), then upon receiving signal $b_{i}$ and
reporting $r_{i}$ and then observing the posted $\pi \in \Pi _{i}(r_{i}),$
buyer $i$ could first compute $(u_{j}(\pi ),c_{j}(\pi )_{j\neq i}$ and then
compute the best response given beliefs at this information set. That is, $i$
could then compute $(w_{i}(\theta ),d_{i}(\theta ))_{\theta \in \Theta }$
that maximizes 
\begin{equation*}
\sum_{b_{-i}\in B_{-i}}\sum_{\theta \in \Theta }\left[ g_{i}(\hat{\varphi}%
(\gamma _{-i}(\pi ,b_{-i}),w_{i}(\pi ),d_{i}(\pi ));u_{i}(\theta
),c_{i}(\theta ),v_{i}(\pi )\right] +\hat{x}_{i}(\gamma _{-i}(\pi
,b_{-i}),w_{i}(\pi ),d_{i}(\pi )))\mu _{i}(\theta ,b_{-i}|r_{i},\pi ,b_{i})
\end{equation*}%
for $\pi \in \Pi _{i}(r_{i}).$ Note that, even if $i$ could compute $%
(u_{j}(\pi ),c_{j}(\pi )_{j\neq i},$ it need not be true that $(u_{i}(\theta
),c_{i}(\theta ))_{\theta \in \Theta }$ is buyer $i$'s best response.

In this paper, we are assuming that $i$ does not know the payoff-externality
profile of the other players nor are we assuming that $i$ has beliefs
regarding these profiles. Along the equilibrium path of play however, buyer $%
i$ need not know the true values of $(u_{j}(\theta ),c_{j}(\theta ))_{j\neq
i}.$ As long as all buyers honestly report their first-stage signals, buyer $%
i$ can determine that he should truthfully report $(u_{i}(\theta
),c_{i}(\theta ))_{\theta \in \Theta }$ in the second stage for any
second-stage strategies of $i$'s opponents.

\section{The problem for large n}

The rewards that ensure honest first-stage reporting by buyers may be quite
large so it is useful to identify a model in which the individual stage 1
rewards, or even the sum of these rewards, is small. To that end, we define
a special sequence of assignment problems $(u^{n},v^{n},c^{n},P^{n})$ with
interdependent valuations. Assume that there exist a finite set $X$ such
that $B_{i}=X$ for all $i.$ Consequently, we write $B=X^{n}$ where $X^{n}$
is the Cartesian product of $n$ copies of $X$. For each $n$ and each $i\in
N=\{1,..,n\},$ let $u_{ij}^{n}(\theta ),c_{ij}^{n,p,q}(\theta )$ and $%
v_{ij}^{n}(\theta )$ denote the valuations, externalities and seller costs
when $i$ is matched with $j$ in state $\theta .$ Furthermore, assume that
for all $n$ and for all $i,j\in N$, $u_{ij}^{n}(\theta )\leq
M,c_{ij}^{n,p,q}(\theta )$ $\leq M$ and $v_{ij}^{n}(\theta )\leq M$ for some 
$M>0.$\newline

\textbf{Definition 3}: A sequence of triples $(u^{n},c^{n},v^{n},P^{n})$ is
a \emph{conditionally independent sequence }of assignment problems with
interdependent valuations if the sequence $P^{n}\in \Delta _{\Theta \times
X^{n}}^{\ast }$ satisfies:

(i) There exists $\lambda \in \Delta ^{\ast }(\Theta )$ and for each $\theta
\in \Theta ,$ there exist a $Q(\cdot |\theta )\in \Delta ^{\ast }(X)$ such
that for each $b\in X^{n},$ 
\begin{equation*}
P^{n}(\theta ,b)=\left[ \prod\limits_{i=1}^{n}Q(b_{i}|\theta )\right]
\lambda (\theta ).
\end{equation*}

(ii) Let $\hat{P}$ denote the common marginal of $P^{n}$ on $B_{i}\times
B_{j}(=X\times X)$ for $i\neq j.$ For each pair $(i,j)$ and $%
b_{i},b_{i}^{\prime }$ in $X$ with $b_{i}\neq b_{i}^{\prime },$ there exists 
$b_{j}\in X$ such that $\hat{P}(b_{j}|b_{i})\neq \hat{P}(b_{j}|b_{i}^{\prime
}).$

\bigskip

\textbf{Theorem 2:} Let $(u^{n},c^{n},v^{n},P^{n})$ be a sequence of
conditionally independent matching problems. There exists an $\hat{n}$ such
that, for all $n>\hat{n}$, there exists a reward system $\xi ^{n}=(\xi
_{i}^{n})_{i\in N}$ such that the two-stage game $\Gamma
(u^{n},v^{n},c^{n},P^{n},\xi ^{n})$ admits a sequential dominant strategy
assessment. Furthermore, for each $k$ we may choose $\xi ^{n}$ so that $%
\sum_{i=1}^{n}\xi _{i}^{n}(b)\sim O\left( \frac{1}{n^{k}}\right) $ for every 
$b\in X^{n}.$

\section{Proof of Theorem 1}

\subsection{Preparatory Lemmas}

\textbf{Lemma A:} Let $M$ be a positive number and let $w_{ij}(\theta
),v_{ij}(\theta ),d_{ij}^{pq}(\theta )$ be a collection of numbers
satisfying $0\leq w_{ij}(\theta ),v_{ij}(\theta ),d_{ij}^{pq}(\theta )\leq M$
for all $i,j,p,q,\theta $. For each $\pi \in \Delta (\Theta ),$ define $%
w_{i}(\pi ),v_{i}(\pi ),d_{i}(\pi )$ as in Section 1.1. For each $\pi \in
\Delta (\Theta ),$ let 
\begin{equation*}
F(\pi )=\max_{z\in Z}\sum_{i=1}^{n}g_{i}(z;w_{i}(\pi ),d_{i}(\pi ),v_{i}(\pi
)).
\end{equation*}%
Then for each $\pi ,\pi ^{\prime }\in \Delta (\Theta ),$

\begin{equation*}
|F(\pi )-F(\pi ^{\prime })|\leq 2nM||\pi -\pi ^{\prime }||.
\end{equation*}

\textbf{Proof}: For each $z\in Z,$ let%
\begin{equation*}
G_{\pi }(z)=\sum_{i=1}^{n}\left[ \sum_{\theta \in \Theta
}g_{i}(z;w_{i}(\theta ),d_{i}(\theta ),v_{i}(\theta ))\pi (\theta )\right]
\end{equation*}%
and let 
\begin{equation*}
\xi (\pi )\in \arg \max_{z\in Z}G_{\pi }(z).
\end{equation*}%
Then 
\begin{eqnarray*}
F(\pi )-F(\pi ^{\prime }) &=&G_{\pi }(\xi (\pi ))-G_{\pi ^{\prime }}(\xi
(\pi ^{\prime })) \\
&=&G_{\pi }(\xi (\pi ))-G_{\pi ^{\prime }}(\xi (\pi ^{\prime }))+G_{\pi
^{\prime }}(\xi (\pi ))-G_{\pi ^{\prime }}(\xi (\pi )) \\
&\leq &G_{\pi }(\xi (\pi ))-G_{\pi ^{\prime }}(\xi (\pi )) \\
&=&\sum_{\theta \in \Theta }\left[ \sum_{i=1}^{n}g_{i}(\xi (\pi
);w_{i}(\theta ),d_{i}(\theta ),v_{i}(\theta ))\right] [\pi (\theta )-\pi
^{\prime }(\theta )] \\
&\leq &(2+n)nM||\pi -\pi ^{\prime }||.
\end{eqnarray*}%
The result follows by reversing the roles of $\pi $ and $\pi ^{\prime }.$

\bigskip

\bigskip \textbf{Lemma B}: Let $M$ be a positive number and let $%
w_{ij}(\theta ),v_{ij}(\theta ),d_{ij}^{pq}(\theta )$ be a collection of
numbers satisfying $0\leq w_{ij}(\theta ),v_{ij}(\theta ),d_{ij}^{pq}(\theta
)\leq M$ for all $i,j,p,q,\theta $. For each $\pi \in \Delta (\Theta ),$
define $w_{i}(\pi ),v_{i}(\pi ),d_{i}(\pi )$ as in Section 1.1. Define

\begin{equation*}
\xi (\pi )\in \arg \max_{z\in Z}\left\{ \sum_{i=1}^{n}g_{i}(z;w_{i}(\pi
),d_{i}(\pi ),v_{i}(\pi ))\right\}
\end{equation*}%
and VCG transfers $\eta _{i}(\pi )$ for each buyer, i.e.,%
\begin{equation*}
\eta _{i}(\pi )=\sum_{\substack{ k  \\ :k\neq i}}g_{k}(\xi (\pi );w_{k}(\pi
),d_{k}(\pi ),v_{k}(\pi ))-\max_{z\in Z_{-i}}\sum_{\substack{ k  \\ :k\neq i 
}}g_{k}(z;w_{k}(\pi ),d_{k}(\pi ),v_{k}(\pi )).
\end{equation*}%
Then for all $\pi ,\pi ^{\prime }\in \Delta (\Theta ),$%
\begin{equation*}
|g_{i}(\xi (\pi ^{\prime });w_{i}(\pi ^{\prime }),d_{i}(\pi ^{\prime
}),v_{i}(\pi ^{\prime }))+\eta _{i}(\pi ^{\prime })-[g_{i}(\xi (\pi
);w_{i}(\pi ),d_{i}(\pi ),v_{i}(\pi ))+\eta _{i}(\pi )]|\leq (2+n)2nM||\pi
-\pi ^{\prime }||.
\end{equation*}

\textbf{Proof}: Applying Lemma A, it follows that 
\begin{eqnarray*}
&&g_{i}(\xi (\pi ^{\prime });w_{i}(\pi ^{\prime }),d_{i}(\pi ^{\prime
}),v_{i}(\pi ^{\prime }))+\eta _{i}(\pi ^{\prime })-[g_{i}(\xi (\pi
);w_{i}(\pi ),d_{i}(\pi ),v_{i}(\pi ))+\eta _{i}(\pi )] \\
&=&\max_{z\in Z}\left[ \sum_{i=1}^{n}g_{i}(z;w_{i}(\pi ^{\prime }),d_{i}(\pi
^{\prime }),v_{i}(\pi ^{\prime }))\right] -\max_{z\in Z}\left[
\sum_{i=1}^{n}g_{i}(z;w_{i}(\pi ),d_{i}(\pi ),v_{i}(\pi ))\right] \\
&&+\max_{z\in Z_{-i}}\left[ \sum_{\substack{ k  \\ :k\neq i}}%
g_{k}(z;w_{k}(\pi ),d_{k}(\pi ),v_{k}(\pi ))\right] -\max_{z\in Z_{-i}}\left[
\sum_{\substack{ k  \\ :k\neq i}}g_{k}(z;w_{k}(\pi ^{\prime }),d_{k}(\pi
^{\prime }),v_{k}(\pi ^{\prime }))\right] \\
&\leq &(2+n)nM||\pi -\pi ^{\prime }||+(2+n)nM||\pi -\pi ^{\prime }||.
\end{eqnarray*}

\subsection{Proof of Theorem 1}

Let $\alpha _{i}^{\ast }(b_{i})=b_{i}$ for each $i$ and $b_{i}$ and recall
that $\beta _{i}^{\ast }$ is defined for buyer $i$ as follows: \ $\beta
_{i}(r_{i},\pi ,b_{i})=(u_{i}(\pi ),c_{i}(\pi ))$ for each $(r_{i},\pi
,b_{i})$ with $r_{i}\in B_{i}.\pi \in \Pi _{i}(r_{i}),$ and $b_{i}\in B_{i}.$

Define beliefs $\mu _{i}^{\ast }(\cdot |r_{i},\pi ,b_{i})\in \Delta (\Theta
\times B_{-i})$ for agent $i$ at each information set $(r_{i},\pi ,b_{i})$
as in Section 2.4 so that, along the equilibrium path of play, we have 
\begin{eqnarray*}
\mu _{i}(\theta ,b_{-i}|b_{i},\pi ,b_{i}) &=&\frac{P_{\Theta }(\theta
|b_{-i},b_{i})P_{-i}(b_{-i}|b_{i})}{\sum_{\hat{b}_{-i}:\rho (\hat{b}%
_{-i},b_{i})=\pi }P_{-i}(\hat{b}_{-i}|b_{i})}\text{ if }\rho
(b_{-i},b_{i})=\pi \\
&=&0\text{ otherwise.}
\end{eqnarray*}

Next, let $\xi _{i}$ be first-stage rewards defined by spherical scoring
rules: for each $b$ and $i$, let 
\begin{equation*}
\xi _{i}(b_{-i},b_{i})=\delta \frac{P_{-i}(b_{-i}|b_{i})}{||P_{-i}(\cdot
|b_{i})||_{2}}.
\end{equation*}%
We will show that $\delta $ can be chosen so that $((\alpha ^{\ast },\beta
^{\ast }),\mu ^{\ast })$ is a sequential dominant strategy assessment in the
game $\Gamma (u,c,v,P,\xi ).$

\bigskip

\emph{Part 1}: The strategy profile $(\alpha ^{\ast },\beta ^{\ast })$ is
truthful. To show that part (ii) of the definition of sequential dominant
strategy assessment is satisfied, suppose that $(\beta _{j})_{j\neq i}$ is a
collection of second-stage strategies of i's opponents. We first lighten the
notation and for each $b_{-i}\in B_{-i}$ and each $k\neq i,$ will write 
\begin{equation*}
\sum_{\theta }[\beta _{k}(b_{k},\pi ,b_{k})](\theta )\pi (\theta )=\gamma
_{k}(\pi ,b_{k})
\end{equation*}%
and 
\begin{equation*}
(\gamma _{j}(\pi ,b_{j}))_{j\neq i}=\gamma _{-i}(\pi ,b_{-i}).
\end{equation*}

We are assuming that all buyers are truthful in stage 1 and we must show
that for each $b_{i}\in B_{i}$ and $\pi \in \Pi (b_{i})$ and each $%
(w_{i}(\theta ),d_{i}(\theta ))_{\theta \in \Theta },$ we have 
\begin{eqnarray*}
&&\sum_{b_{-i}\in B_{-i}}\sum_{\theta \in \Theta }{\LARGE [}g_{i}(\hat{%
\varphi}(\gamma _{-i}(\pi ,b_{-i}),u_{i}(\pi ),c_{i}(\pi ));u_{i}(\theta
),c_{i}(\theta ),v_{i}(\pi ))+\hat{x}_{i}(\gamma _{-i}(\pi
,b_{-i}),u_{i}(\pi ),c_{i}(\pi ))){\LARGE ]}\mu _{i}(\theta
,b_{-i}|b_{i},\pi ,b_{i}) \\
&\geq &\sum_{b_{-i}\in B_{-i}}\sum_{\theta \in \Theta }{\LARGE [}g_{i}(\hat{%
\varphi}(\gamma _{-i}(\pi ,b_{-i}),w_{i}(\pi ),d_{i}(\pi ));u_{i}(\theta
),c_{i}(\theta ),v_{i}(\pi ))+\hat{x}_{i}(\gamma _{-i}(\pi
,b_{-i}),w_{i}(\pi ),d_{i}(\pi )){\Large ]}\mu _{i}(\theta ,b_{-i}|b_{i},\pi
,b_{i}).
\end{eqnarray*}

To see this, first note that the dominant strategy property of the mechanism
implies that for each $\pi $ and $b_{-i},$ 
\begin{eqnarray*}
&&g_{i}(\hat{\varphi}(\gamma _{-i}(\pi ,b_{-i}),w_{i}(\pi ),d_{i}(\pi
));u_{i}(\pi ),c_{i}(\pi ),v_{i}(\pi ))+x_{i}(\gamma _{-i}(\pi
,b_{-i}),w_{i}(\pi ),d_{i}(\pi )) \\
&\leq &g_{i}(\hat{\varphi}(\gamma _{-i}(\pi ,b_{-i}),u_{i}(\pi ),c_{i}(\pi
));u_{i}(\pi ),c_{i}(\pi ),v_{i}(\pi ))+x_{i}(\gamma _{-i}(\pi
,b_{-i}),u_{i}(\pi ),c_{i}(\pi )).
\end{eqnarray*}%
Consequently, 
\begin{eqnarray*}
&&\text{{\footnotesize $\sum_{b_{-i}\in B_{-i}}\sum_{\theta \in \Theta }%
\Biggr[g_{i}(\hat{\varphi}(\gamma_{-i}(\pi ,b_{-i}),w_{i}(\pi ),d_{i}(\pi
));u_{i}(\theta ),c_{i}(\theta ),v_{i}(\pi ))+\hat{x}_{i}(\gamma _{-i}(\pi
,b_{-i}),w_{i}(\pi ),d_{i}(\pi ))\Biggr]\mu _{i}(\theta ,b_{-i}|b_{i},\pi
,b_{i})$}} \\
&\text{{\footnotesize $=$}}&\text{{\footnotesize $\sum_{\substack{ b_{-i}\in
B_{-i}  \\ :\rho (b_{-i},b_{i})=\pi }}\sum_{\theta \in \Theta }\Biggr[g_{i}(%
\hat{\varphi}(\gamma _{-i}(\pi ,b_{-i}),w_{i}(\pi ),d_{i}(\pi
));u_{i}(\theta ),c_{i}(\theta ),v_{i}(\pi ))+ \hat{x}_{i}(\gamma _{-i}(\pi
,b_{-i}),w_{i}(\pi ),d_{i}(\pi ))\Biggr]\frac{P_{\Theta }(\theta
|b_{-i},b_{i})P_{-i}(b_{-i}|b_{i})}{\smashoperator{\sum_{\hat{b}_{-i}:\rho
(\hat{b}_{-i},b_{i})=\pi }}P_{-i}(\hat{b}_{-i}b_{i})} $}} \\
&\text{{\footnotesize $=$}}&\text{{\footnotesize $\sum_{\substack{ b_{-i}\in
B_{-i}  \\ :\rho (b_{-i},b_{i})=\pi }}\left[ \sum_{\theta \in \Theta }g_{i}(%
\hat{\varphi}(\gamma _{-i}(\pi ,b_{-i}),w_{i}(\pi ),d_{i}(\pi
));u_{i}(\theta ),c_{i}(\theta ),v_{i}(\pi ))P_{\Theta }(\theta
|b_{-i},b_{i}) +\hat{x}_{i}(\gamma _{-i}(\pi ,b_{-i}),w_{i}(\pi ),d_{i}(\pi
))\right] \frac{P_{-i}(b_{-i}|b_{i})}{\smashoperator{\sum_{\hat{b}_{-i}:\rho
(\hat{b}_{-i},b_{i})=\pi }}P_{-i}(\hat{b}_{-i}b_{i})}$}} \\
&\text{{\footnotesize $=$}}&\text{{\footnotesize $\sum_{\substack{ b_{-i}\in
B_{-i}  \\ :\rho (b_{-i},b_{i})=\pi }}\Biggr[ g_{i}(\hat{\varphi}(\gamma
_{-i}(\pi ,b_{-i}),w_{i}(\pi ),d_{i}(\pi ));u_{i}(\pi ),c_{i}(\pi
),v_{i}(\pi ))+\hat{x}_{i}(\gamma _{-i}(\pi ,b_{-i}),w_{i}(\pi ),d_{i}(\pi ))%
\Biggr] \frac{P_{-i}(b_{-i}|b_{i})}{\smashoperator{\sum_{\hat{b}_{-i}:\rho
(\hat{b}_{-i},b_{i})=\pi }}P_{-i}(\hat{b}_{-i}b_{i})}$}} \\
&\text{{\footnotesize $\leq$}}&\text{{\footnotesize $\sum_{\substack{ %
b_{-i}\in B_{-i}  \\ :\rho (b_{-i},b_{i})=\pi }}\Biggr[ g_{i}(\hat{\varphi}%
(\gamma _{-i}(\pi ,b_{-i}),u_{i}(\pi ),c_{i}(\pi ));u_{i}(\pi ),c_{i}(\pi
),v_{i}(\pi )) +\hat{x}_{i}(\gamma _{-i}(\pi ,b_{-i}),u_{i}(\pi ),c_{i}(\pi
))\Biggr]\frac{P_{-i}(b_{-i}|b_{i})}{\smashoperator{\sum_{\hat{b}_{-i}:\rho
(\hat{b}_{-i},b_{i})=\pi }}P_{-i}(\hat{b}_{-i}b_{i})}$}} \\
&\text{{\footnotesize $=$}}&\text{{\footnotesize $\sum_{b_{-i}\in
B_{-i}}\sum_{\theta \in \Theta }[g_{i}(\hat{\varphi}(\gamma _{-i}(\pi
,b_{-i}),u_{i}(\pi ),c_{i}(\pi );u_{i}(\theta ),c_{i}(\theta ),v_{i}(\pi ))+%
\hat{x}_{i}(\gamma _{-i}(\pi ,b_{-i}),u_{i}(\pi ),c_{i}(\pi ))]\mu
_{i}(\theta ,b_{-i}|b_{i},\pi ,b_{i})$}}
\end{eqnarray*}

where the penultimate inequality follows from the dominant strategy property
of the VCG mechanism. \newline

\emph{Part 2}: To complete the proof, we must show that first-stage rewards $%
\xi _{i}$ can be chosen so that coordinated deviations across stages are
unprofitable for buyer $i$ given any behavior strategy profile $(\alpha
_{-i},\beta _{-i})$ of the other players. To see this, choose $%
b=(b_{-i},b_{i}),$ $r_{i}\in B_{i}$ and let $\pi =\rho (b)$ and $\pi
^{\prime }=\rho (b_{-i},r).$ For each $j\neq i$ and $\theta ,$ let 
\begin{equation*}
\sum_{\theta \in \Theta }[\beta _{j}(\alpha _{j}(b_{j}),\pi ,b_{j})](\theta
)\pi (\theta )=\gamma _{j}(\pi ,b_{j})
\end{equation*}%
and 
\begin{equation*}
\sum_{\theta \in \Theta }[\beta _{j}(\alpha _{j}(b_{j}),\pi ^{\prime
},b_{j})](\theta )\pi ^{\prime }(\theta )=\gamma _{j}(\pi ^{\prime },b_{j}).
\end{equation*}%
Let $(\gamma _{j}(\pi ,b_{j}))_{j\neq i}=\gamma _{-i}(\pi ,b_{-i})$ and $%
(\gamma _{j}(\pi ^{\prime },b_{j}))_{j\neq i}=\gamma _{-i}(\pi ^{\prime
},b_{-i}).$ We claim that, for all $(w_{i}(\theta ),d_{i}(\theta ))_{\theta
\in \Theta },$ we have%
\begin{eqnarray*}
&&g_{i}(\hat{\varphi}(\gamma _{-i}(\pi ,b_{-i}),u_{i}(\pi ),c_{i}(\pi
));u_{i}(\pi ),c_{i}(\pi ),v_{i}(\pi ))+\hat{x}_{i}(\gamma _{-i}(\pi
,b_{-i}),u_{i}(\pi ),c_{i}(\pi )) \\
&&-[g_{i}(\hat{\varphi}(\gamma _{-i}(\pi ^{\prime },b_{-i}),w_{i}(\pi
^{\prime }),d_{i}(\pi ^{\prime }));u_{i}(\pi ),c_{i}(\pi ),v_{i}(\pi
^{\prime }))+\hat{x}_{i}(\gamma _{-i}(\pi ^{\prime },b_{-i}),w_{i}(\pi
^{\prime }),d_{i}(\pi ^{\prime }))] \\
&\geq &-(2+n)(2n+1)M||\pi -\pi ^{\prime }||.
\end{eqnarray*}

First, note that 
\begin{eqnarray*}
&&|g_{i}(\hat{\varphi}(\gamma _{-i}(\pi ^{\prime },b_{-i}),w_{i}(\pi
^{\prime }),d_{i}(\pi ^{\prime }));u_{i}(\pi ^{\prime }),c_{i}(\pi ^{\prime
}),v_{i}(\pi ^{\prime }))-[g_{i}(\hat{\varphi}(\gamma _{-i}(\pi ^{\prime
},b_{-i}),w_{i}(\pi ^{\prime }),d_{i}(\pi ^{\prime }));u_{i}(\pi ),c_{i}(\pi
),v_{i}(\pi ))]| \\
&=&|\sum_{\theta }g_{i}(\hat{\varphi}(\gamma _{-i}(\pi ^{\prime
},b_{-i}),w_{i}(\pi ^{\prime }),d_{i}(\pi ^{\prime }));u_{i}(\theta
),c_{i}(\theta ),v_{i}(\theta ))[\pi ^{\prime }(\theta )-\pi (\theta )]| \\
&\leq &(2+n)M||\pi ^{\prime }-\pi ||
\end{eqnarray*}%
and 
\begin{eqnarray*}
&&|g_{i}(\hat{\varphi}(\gamma _{-i}(\pi ^{\prime },b_{-i}),w_{i}(\pi
^{\prime }),d_{i}(\pi ^{\prime }));u_{i}(\pi ),c_{i}(\pi ),v_{i}(\pi
))-[g_{i}(\hat{\varphi}(\gamma _{-i}(\pi ^{\prime },b_{-i}),w_{i}(\pi
^{\prime }),d_{i}(\pi ^{\prime }));u_{i}(\pi ),c_{i}(\pi ),v_{i}(\pi
^{\prime }))]| \\
&=&|\sum_{\theta }g_{i}(\hat{\varphi}(\gamma _{-i}(\pi ^{\prime
},b_{-i}),w_{i}(\pi ^{\prime }),d_{i}(\pi ^{\prime }));u_{i}(\pi ),c_{i}(\pi
),v_{i}(\theta ))[\pi (\theta )-\pi ^{\prime }(\theta )]| \\
&\leq &(2+n)M||\pi ^{\prime }-\pi ||.
\end{eqnarray*}

Since the dominant strategy property of the VCG mechanism implies that

\begin{eqnarray*}
&&g_{i}(\hat{\varphi}(\gamma _{-i}(\pi ^{\prime },b_{-i}),u_{i}(\pi ^{\prime
}),c_{i}(\pi ^{\prime }));u_{i}(\pi ^{\prime }),c_{i}(\pi ^{\prime
}),v_{i}(\pi ^{\prime }))+\hat{x}_{i}(w_{j}^{\prime }(\pi ),d_{j}^{\prime
}(\pi ))_{j\neq i},u_{i}(\pi ^{\prime }),c_{i}(\pi ^{\prime })) \\
&\geq &g_{i}(\hat{\varphi}(\gamma _{-i}(\pi ^{\prime },b_{-i}),w_{i}(\pi
^{\prime }),d_{i}(\pi ^{\prime });u_{i}(\pi ^{\prime }),c_{i}(\pi ^{\prime
}),v_{i}(\pi ^{\prime })+\hat{x}_{i}((w_{j}^{\prime }(\pi ),d_{j}^{\prime
}(\pi ))_{j\neq i},w_{i}(\pi ^{\prime }),d_{i}(\pi ^{\prime }))
\end{eqnarray*}

it then follows from Lemmas A and B that%
\begin{eqnarray*}
&&g_{i}(\hat{\varphi}(\gamma _{-i}(\pi ,b_{-i}),u_{i}(\pi ),c_{i}(\pi
));u_{i}(\pi ),c_{i}(\pi ),v_{i}(\pi ))+\hat{x}_{i}(\gamma _{-i}(\pi
,b_{-i}),u_{i}(\pi ),c_{i}(\pi )) \\
&&-[g_{i}(\hat{\varphi}(\gamma _{-i}(\pi ^{\prime },b_{-i}),w_{i}(\pi
^{\prime }),d_{i}(\pi ^{\prime }));u_{i}(\pi ),c_{i}(\pi ),v_{i}(\pi
^{\prime }))+\hat{x}_{i}(\gamma _{-i}(\pi ^{\prime },b_{-i}),w_{i}(\pi
^{\prime }),d_{i}(\pi ^{\prime }))] \\
&=&g_{i}(\hat{\varphi}(\gamma _{-i}(\pi ,b_{-i}),u_{i}(\pi ),c_{i}(\pi
));u_{i}(\pi ),c_{i}(\pi ),v_{i}(\pi ))+\hat{x}_{i}(\gamma _{-i}(\pi
,b_{-i}),u_{i}(\pi ),c_{i}(\pi )) \\
&&-[g_{i}(\hat{\varphi}(\gamma _{-i}(\pi ^{\prime },b_{-i}),u_{i}(\pi
^{\prime }),c_{i}(\pi ^{\prime }));u_{i}(\pi ^{\prime }),c_{i}(\pi ^{\prime
}),v_{i}(\pi ^{\prime }))+\hat{x}_{i}(\gamma _{-i}(\pi ^{\prime
},b_{-i}),u_{i}(\pi ^{\prime }),c_{i}(\pi ^{\prime }))] \\
&&+g_{i}(\hat{\varphi}(\gamma _{-i}(\pi ^{\prime },b_{-i}),u_{i}(\pi
^{\prime }),c_{i}(\pi ^{\prime }));u_{i}(\pi ^{\prime }),c_{i}(\pi ^{\prime
}),v_{i}(\pi ^{\prime }))+\hat{x}_{i}(\gamma _{-i}(\pi ^{\prime
},b_{-i}),u_{i}(\pi ^{\prime }),c_{i}(\pi ^{\prime })) \\
&&-[g_{i}(\hat{\varphi}(\gamma _{-i}(\pi ^{\prime },b_{-i}),w_{i}(\pi
^{\prime }),d_{i}(\pi ^{\prime }));u_{i}(\pi ^{\prime }),c_{i}(\pi ^{\prime
}),v_{i}(\pi ^{\prime }))+\hat{x}_{i}(\gamma _{-i}(\pi ^{\prime
},b_{-i}),w_{i}(\pi ^{\prime }),d_{i}(\pi ^{\prime }))] \\
&&+g_{i}(\hat{\varphi}(\gamma _{-i}(\pi ^{\prime },b_{-i}),w_{i}(\pi
^{\prime }),d_{i}(\pi ^{\prime }));u_{i}(\pi ^{\prime }),c_{i}(\pi ^{\prime
}),v_{i}(\pi ^{\prime })) \\
&&-g_{i}(\hat{\varphi}(\gamma _{-i}(\pi ^{\prime },b_{-i}),w_{i}(\pi
^{\prime }),d_{i}(\pi ^{\prime }));u_{i}(\pi ),c_{i}(\pi ),v_{i}(\pi )) \\
&&+g_{i}(\hat{\varphi}(\gamma _{-i}(\pi ^{\prime },b_{-i}),w_{i}(\pi
^{\prime }),d_{i}(\pi ^{\prime }));u_{i}(\pi ),c_{i}(\pi ),v_{i}(\pi )) \\
&&-g_{i}(\hat{\varphi}(\gamma _{-i}(\pi ^{\prime },b_{-i}),w_{i}(\pi
^{\prime }),d_{i}(\pi ^{\prime }));u_{i}(\pi ),c_{i}(\pi ),v_{i}(\pi
^{\prime })) \\
&\geq &-[(2+n)2nM||\pi -\pi ^{\prime }||+(2+n)M||\pi ^{\prime }-\pi
||+(2+n)M||\pi ^{\prime }-\pi ||] \\
&=&-(2+n)(2n+1)M||\pi -\pi ^{\prime }||.
\end{eqnarray*}

\emph{Part 3}: To complete the proof, we must show that the first-stage
rewards $\xi _{i}$ can be chosen so that for all $r_{i}\in B_{i}$ and all $%
(w_{i}(\theta ),d_{i}(\theta ))_{\theta \in \Theta },$ we have \ 
\begin{eqnarray*}
&&\sum_{b_{-i}\in B_{-i}}[g_{i}(\hat{\varphi}(\gamma _{-i}(\rho
(b),b_{-i}),u_{i}(\rho (b)),c_{i}(\rho (b));u_{i}(\rho (b)),c_{i}(\rho
(b)),v_{i}(\rho (b)))+ \\
&&+\hat{x}_{i}(\gamma _{-i}(\rho (b),b_{-i}),u_{i}(\rho (b)),c_{i}(\rho
(b)))+\xi _{i}(b)]P(b_{-i}|b_{i})\geq \\
&\geq &\sum_{b_{-i}\in B_{-i}}[g_{i}(\hat{\varphi}(\gamma _{-i}(\rho
(b_{-i},r_{i}),b_{-i}),w_{i}(\rho (b_{-i},r_{i})),d_{i}(\rho
(b_{-i},r_{i}));u_{i}(\rho (b)),c_{i}(\rho (b)),v_{i}(\rho (b_{-i},r_{i})))+
\\
&&+\hat{x}_{i}(\gamma _{-i}(\rho (b_{-i},r_{i}),b_{-i}),w_{i}(\rho
(b_{-i},r_{i})),d_{i}(\rho (b_{-i},r_{i})))+\xi
_{i}(b_{-i},r_{i})]P(b_{-i}|b_{i})
\end{eqnarray*}

If $\delta >0$ and 
\begin{equation*}
\xi _{i}(b_{-i},b_{i})=\delta \frac{P_{-i}(b_{-i}|b_{i})}{||P_{-i}(\cdot
|b_{i})||_{2}}
\end{equation*}%
then 
\begin{equation*}
\sum_{(b_{-i},s)}{\LARGE [}\xi _{i}(b_{-i},b_{i})-\xi _{i}(b_{-i},r_{i})%
{\LARGE ]}P_{-i}(b_{-i}|b_{i}))>0
\end{equation*}%
whenever $b_{i}\neq r_{i}$ (recall that $P(\cdot |b_{i})\neq P(\cdot |r_{i})$
whenever $b_{i}\neq r_{i}).$ Therefore, we can choose $\delta >0$ so that%
\begin{eqnarray*}
&&\sum_{b_{-i}\in B_{-i}}\Biggr[g_{i}(\hat{\varphi}(\beta _{j}(\alpha
_{j}(b_{j}),\rho (b),b_{j})_{j\neq i},u_{i}(\rho (b)),c_{i}(\rho
(b)));u_{i}(\rho (b)),c_{i}(\rho (b)),v_{i}(\rho (b)))+ \\
&&+\hat{x}_{i}(\beta _{j}(\alpha _{j}(b_{j}),\rho (b),b_{j})_{j\neq
i},u_{i}(\rho (b)),c_{i}(\rho (b)))+\xi _{i}(b)\Biggr]P(b_{-i}|b_{i})- \\
&&-\Biggr[\sum_{b_{-i}\in B_{-i}}[g_{i}(\hat{\varphi}(\beta _{j}(\alpha
_{j}(b_{j}),\rho (b_{-i},r_{i}),b_{j})_{j\neq i},w_{i},d_{i});u_{i}(\rho
(b)),c_{i}(\rho (b)),v_{i}(\rho (b_{-i},r_{i})))+ \\
&&+\hat{x}_{i}(\beta _{j}(\alpha _{j}(b_{j}),\rho
(b_{-i},r_{i}),b_{j})_{j\neq i},w_{i},d_{i})+\xi
_{i}(b_{-i},r_{i})]P(b_{-i}|b_{i})\Biggr]\geq \\
&\geq &-(2+n)(2n+1)M\sum_{b_{-i}}||\rho (b)-\rho
(b_{-i},r_{i})||P_{-i}(b_{-i}|b_{i})+\sum_{b_{-i}}{\LARGE [}\xi _{i}(b)-\xi
_{i}(b_{-i},r_{i}){\LARGE ]}P_{-i}(b_{-i}|b_{i}) \\
&>&0.
\end{eqnarray*}

This completes the proof of Theorem 1.

\section{Proof of Theorem 2}

\subsection{Preliminaries}

The proof of Theorem 2 is a close adaptation of the proof of Theorem 3 in 
\cite{mclean2015implementation} so we omit most of the details. As in that
proof, we will treat $P^{n}\in \Delta _{\Theta \times X^{n}}^{\ast }$ as the
distribution of a $(\Theta \times X^{n})$-valued random variable which we
denote $(\tilde{\theta},\tilde{b}),$ i.e., 
\begin{equation*}
\text{ Prob}\{(\tilde{\theta},\tilde{b})=(\theta ,b)\}=P^{n}(\theta ,b)
\end{equation*}%
for each $(\theta ,b)\in \Theta \times X^{n}.$ For each $n$ and each $i$ and 
$j$, define 
\begin{equation*}
\nu _{i}^{n}=\max_{b_{i}\in X}\max_{r_{i}\in X}\min \{\varepsilon \geq 0|||%
\text{Prob}\{P_{\Theta }^{n}(\cdot |\tilde{b}_{-i},b_{i})-P_{\Theta
}^{n}(\cdot |\tilde{b}_{-i},r_{i})||>\varepsilon |\tilde{b}_{i}=b_{i}\}\leq
\varepsilon \}
\end{equation*}

From the proof of Lemma B in ~\cite{mclean2015implementation} we conclude
that $\nu _{i}^{n}\sim O\left( \frac{1}{n^{k}}\right) $ for every positive
integer $k$.

Next, suppose that $b=(b_{-i},b_{i})\in X^{n}$ and $r_{i}\in X.$ Define $%
\rho ^{n}(b)=P_{\Theta }^{n}(\cdot |b)$ and $\rho
^{n}(b_{-i},r_{i})=P_{\Theta }^{n}(\cdot |b_{-i},r_{i}).$ Let $M$ be the
bound defined in the statement of Theorem 2. From step 2 in the proof of
Theorem 1, it follows that

\begin{eqnarray*}
&&g_{i}(\hat{\varphi}(\gamma _{-i}(\rho ^{n}(b),b_{-i}),u_{i}^{n}(\rho
^{n}(b)),c_{i}^{n}(\rho ^{n}(b)));u_{i}^{n}(\rho ^{n}(b)),c_{i}^{n}(\rho
^{n}(b)),v_{i}^{n}(\rho ^{n}(b)))+\hat{x}_{i}(\gamma _{-i}(\rho
^{n}(b),b_{-i}),u_{i}^{n}(\rho ^{n}(b)),c_{i}^{n}(\rho ^{n}(b))) \\
&&-[g_{i}(\hat{\varphi}(\gamma _{-i}(\rho
^{n}(b_{-i},r_{i}),b_{-i}),w_{i}(\rho ^{n}(b_{-i},r_{i})),d_{i}(\rho
^{n}(b_{-i},r_{i})));u_{i}^{n}(\rho ^{n}(b)),c_{i}^{n}(\rho
^{n}(b)),v_{i}^{n}(\rho ^{n}(b_{-i},r_{i})))+ \\
&&+\hat{x}_{i}(\gamma _{-i}(\rho ^{n}(b_{-i},r_{i}),b_{-i}),w_{i}(\rho
^{n}(b_{-i},r_{i})),d_{i}(\rho ^{n}(b_{-i},r_{i})))]\geq -(2+n)(2n+1)M||\rho
^{n}(b)-\rho ^{n}(b_{-i},r_{i})||.
\end{eqnarray*}%
To prove Theorem 2, it suffices to show that we can find for each $k$
first-stage rewards $\xi _{i}$ such that%
\begin{equation*}
n^{k}\sum_{i=1}^{n}\xi _{i}(b)\underset{n\rightarrow \infty }{\rightarrow }0
\end{equation*}%
and 
\begin{equation*}
-(2+n)(2n+1)M\sum_{b_{-i}\in X^{n-1}}||\rho ^{n}(b)-\rho
^{n}(b_{-i},r_{i})||P_{-i}^{n}(b_{-i}|b_{i})+\sum_{b_{-i}\in X^{n-1}}{\LARGE %
[}\xi _{i}(b)-\xi _{i}(b_{-i},r_{i}){\LARGE ]}P_{-i}^{n}(b_{-i}|b_{i})>0
\end{equation*}%
for all sufficiently large $n$.

Choose $k>1$ and for each $b=(b_{1},..,b_{n})\in X^{n},$ define

\begin{align*}
\xi _{i}(b_{-i},b_{i})& =\frac{1}{n^{k+2}}\frac{\hat{P}(b_{i+1}|b_{i})}{||%
\hat{P}(\cdot |b_{i})||_{2}}\text{ if }i=1,..,n-1 \\
& =\frac{1}{n^{k+2}}\frac{\hat{P}(b_{1}|b_{i})}{||\hat{P}(\cdot |b_{i})||_{2}%
}\text{ if }i=n
\end{align*}%
where $\hat{P}$ is defined as in Theorem 2. Therefore, 
\begin{equation*}
0\leq \sum_{i=1}^{n}\xi _{i}(b_{-i},b_{i})\leq \frac{1}{n^{k+1}}
\end{equation*}%
for all $i$, $b_{-i}$ and $b_{i}$.

Let $|X|$ denote the cardinality of $X$ and let $K=\frac{|X|^{-\frac{5}{2}}}{%
4(|X|-1)}.$ Applying Lemma A.3 in~\cite{mclean2004informational}, we
conclude that 
\begin{eqnarray*}
&&\sum_{b_{-i}}\left( \xi _{i}(b)-\xi _{i}(b_{-i},r_{i})\right)
P_{-i}^{n}(b_{-i}|b_{i}) \\
&=&\sum_{b_{-i}}\frac{1}{n^{k+2}}\left[ \frac{\hat{P}(b_{i+1}|b_{i})}{||\hat{%
P}^{n}(\cdot |b_{i})||_{2}}-\frac{\hat{P}(b_{i+1}|r_{i})}{||\hat{P}%
^{n}(\cdot |r_{i})||_{2}}\right] P_{-i}^{n}(b_{-i},|b_{i}) \\
&=&\sum_{x\in X}\frac{1}{n^{k+2}}\left[ \frac{\hat{P}(x|b_{i})}{||\hat{P}%
(\cdot |b_{i})||_{2}}-\frac{\hat{P}(x|r_{i})}{||\hat{P}(\cdot |r_{i})||_{2}}%
\right] \hat{P}^{n}(x|b_{i}) \\
&>&\frac{1}{n^{k+2}}K(||\hat{P}(\cdot |b_{i})-\hat{P}(\cdot |r_{i})||)^{2}.
\end{eqnarray*}

To complete the argument, define 
\begin{equation*}
A_{i}(r_{i},b_{i})=\{b_{-i}\in B_{-i}|\text{ }||P_{\Theta }^{n}(\cdot
|b_{-i},b_{i})-P_{\Theta }^{n}(\cdot |b_{-i},r_{i})||>\hat{\nu}_{i}^{n}\}
\end{equation*}%
and 
\begin{equation*}
C_{i}(r_{i},b_{i})=\{b_{-i}\in B_{-i}|\text{ }||P_{\Theta }^{n}(\cdot
|b_{-i},b_{i})-P_{\Theta }^{n}(\cdot |b_{-i},r_{i})||\leq \hat{\nu}%
_{i}^{n}\}.
\end{equation*}

Since%
\begin{equation*}
\text{Prob}\{\tilde{b}_{-i}\in A_{i}(r_{i},b_{i})|\tilde{b}_{i}=b_{i}\}\leq
\nu ^{n}
\end{equation*}%
we conclude that%
\begin{eqnarray*}
&&\sum_{b_{-i},\in X^{n-1}}||\rho ^{n}(b)-\rho
^{n}(b_{-i},r_{i})||P^{n}(b_{-i}|b_{i}) \\
&=&\sum_{b_{-i}\in A_{i}(r_{i},b_{i})}||\rho ^{n}(b)-\rho
^{n}(b_{-i},r_{i})||P^{n}(b_{-i}|b_{i})+\sum_{b_{-i}\in
C_{i}(r_{i},b_{i})}||\rho ^{n}(b)-\rho ^{n}(b_{-i},r)||P^{n}(b_{-i}|b_{i}) \\
&\leq &2\nu ^{n}
\end{eqnarray*}%
Therefore, for all sufficiently large $n$ we have

\begin{eqnarray*}
&&-(2+n)(2n+1)M\sum_{b_{-i}}||\rho ^{n}(b)-\rho
^{n}(b_{-i},r_{i})||P_{-i}^{n}(b_{-i}|b_{i})+\sum_{b_{-i}}{\LARGE [}\xi
_{i}(b)-\xi _{i}(b_{-i},r_{i}){\LARGE ]}P_{-i}^{n}(b_{-i}|b_{i}) \\
&\geq &-(2+n)(2n+1)M2\nu ^{n}+\frac{1}{n^{k+2}}K(||\hat{P}(\cdot |b_{i})-%
\hat{P}(\cdot |r_{i})||)^{2} \\
&=&\frac{1}{n^{k+2}}\left[ K(||\hat{P}(\cdot |b_{i})-\hat{P}(\cdot
|r_{i})||)^{2}-n^{k+2}(2+n)(2n+1)M2\nu ^{n}\right] \\
&>&0.
\end{eqnarray*}

\section{Discussion}

\bigskip

1. We have assumed that the valuations $u_{i}$ and externality costs $c_{i}$
of buyer $i$ are functions only of the unobserved state $\theta .$ At the
cost of more complex notation, we could allow these functions to depend also
on the signal $b_{i}.$

\bigskip

2. In our model, buyers receive private signals that are correlated with the
state $\theta $. Furthermore, the valuations ($u_{i}(\theta ))_{\theta \in
\Theta }$ and externalities ($c_{i}(\theta ))_{\theta \in \Theta }$ of each
buyer $i$ are also that buyer's private information while the costs ($%
v_{i}(\theta ))_{\theta \in \Theta }$ are known to all agents and the
mechanism. As a result, sellers are not strategic actors in our model. We
could allow each seller $j$ to also receive a private signal correlated with
the state. Consequently, the mechanism would now want to elicit the
privately observed signals of both buyers and sellers in order to compute an
efficient allocation. If we retain the assumption that the surplus generated
by each pairing in an optimal assignment goes to the buyer (as we do in this
paper) and assume that the sellers' costs (but not their signals) are known
to all agents and the mechanism, then we\ can find first-stage rewards for
both buyers and sellers that induce honest reporting in the first stage.

\bigskip

3. Our two-stage formulation allows us to resolve the interdependency
resulting from private signals so that the second-stage problem is reduced
to a simpler implementation problem to which the classic private values VCG
transfers can be applied to elicit the buyers' valuations and externalities.
Two natural extensions of the model would pose challenges.

For example, we could allow the surplus generated by each pairing in an
optimal assignment to go to the buyer (as we do in this paper) but assume
that the sellers' valuations, as well as buyers' valuations, to be private
information. The mechanism would then have to elicit the valuations of
buyers and the costs of sellers. In this case, the second stage following
the announcement of first-stage signals would no longer be an implementation
problem with private values since the payoff of buyer $i$ depends on the
valuation of the seller with whom he is matched and that valuation is
private information of the seller. Consequently, the classical private
values VCG transfers do not typically provide incentives for honest
reporting.

\bigskip

4. In this paper we have made weak assumptions concerning what players and
the mechanism know regarding the data of the game. All agents and the
mechanism know the probability measure $P\in \Delta ^{\ast }(\Theta \times
B) $, the sellers' costs $(v_{i}(\theta ))_{\theta \in \Theta },$ and the
bound $M$ on valuations, externalities and costs. However, buyer $i$'s
valuations and externality parameters $(u_{i}(\theta ),c_{i}(\theta
))_{\theta \in \Theta }$ are known only to $i$. As we have emphasized, we do
not take a Bayesian viewpoint and propose beliefs for $i$ regarding the
valuations and externalities of other buyers. There are drawbacks and
advantages to these assumptions. Our model is arguably best tailored to
situations in which n is large for two reasons. First, it is the case of
large $n$ where the assumption of common knowledge of valuations and
externalities seems least plausible. Second, as per Theorem 2, we can find
for every $k$ in the conditionally independent framework a sequence of
first-period rewards ($\xi _{i}^{n})_{i=1}^{n}$ such that $%
n^{k}\sum_{i=1}^{n}\xi _{i}^{n}(b)\rightarrow 0$ as $n\rightarrow \infty .$
Consequently, we obtain a sequential dominant strategy assessment with
asymptotically negligible aggregate first-period payments under a weak
informational assumption. There is a cost associated with this weak
informational assumption, with or without large numbers. At second-stage
information sets through which the equilibrium path does not pass, we cannot
determine if our proposed equilibrium assessment of Theorem 1 is
sequentially rational. If valuations and externalities are common knowledge
and signals provide the only source of asymmetric information (as in~\cite%
{mclean2017dynamic}), then we can in that case propose a truthful
equilibrium that is sequentially rational at every information set thus
yielding a Perfect Bayesian equilibrium.

\bibliography{bib}

\begin{thebibliography}{13}
\providecommand{\natexlab}[1]{#1}
\providecommand{\url}[1]{\texttt{#1}}
\expandafter\ifx\csname urlstyle\endcsname\relax
  \providecommand{\doi}[1]{doi: #1}\else
  \providecommand{\doi}{doi: \begingroup \urlstyle{rm}\Url}\fi

\bibitem[Ausubel(2006)]{ausubel2006efficient}
Lawrence~M Ausubel.
\newblock An efficient dynamic auction for heterogeneous commodities.
\newblock \emph{American Economic Review}, 96\penalty0 (3):\penalty0 602--629,
  2006.

\bibitem[Demange and Gale(1985)]{demange1985strategy}
Gabrielle Demange and David Gale.
\newblock The strategy structure of two-sided matching markets.
\newblock \emph{Econometrica: Journal of the Econometric Society}, pages
  873--888, 1985.

\bibitem[Fernandez et~al.(2022)Fernandez, Rudov, and
  Yariv]{fernandez2022centralized}
Marcelo~Ariel Fernandez, Kirill Rudov, and Leeat Yariv.
\newblock Centralized matching with incomplete information.
\newblock \emph{American Economic Review: Insights}, 4\penalty0 (1):\penalty0
  18--33, 2022.

\bibitem[Jehiel et~al.(1996)Jehiel, Moldovanu, and Stacchetti]{jehiel1996not}
Philippe Jehiel, Benny Moldovanu, and Ennio Stacchetti.
\newblock How (not) to sell nuclear weapons.
\newblock \emph{The American Economic Review}, pages 814--829, 1996.

\bibitem[Klemperer(2004)]{klemperer2004auctions}
Paul Klemperer.
\newblock \emph{Auctions: theory and practice}.
\newblock Princeton University Press, 2004.

\bibitem[Leonard(1983)]{leonard1983elicitation}
Herman~B Leonard.
\newblock Elicitation of honest preferences for the assignment of individuals
  to positions.
\newblock \emph{Journal of political Economy}, 91\penalty0 (3):\penalty0
  461--479, 1983.

\bibitem[Liu et~al.(2014)Liu, Mailath, Postlewaite, and
  Samuelson]{liu2014stable}
Qingmin Liu, George~J Mailath, Andrew Postlewaite, and Larry Samuelson.
\newblock Stable matching with incomplete information.
\newblock \emph{Econometrica}, 82\penalty0 (2):\penalty0 541--587, 2014.

\bibitem[McLean and Postlewaite(2004)]{mclean2004informational}
Richard McLean and Andrew Postlewaite.
\newblock Informational size and efficient auctions.
\newblock \emph{The Review of Economic Studies}, 71\penalty0 (3):\penalty0
  809--827, 2004.

\bibitem[McLean and Postlewaite(2015)]{mclean2015implementation}
Richard~P McLean and Andrew Postlewaite.
\newblock Implementation with interdependent valuations.
\newblock \emph{Theoretical Economics}, 10\penalty0 (3):\penalty0 923--952,
  2015.

\bibitem[McLean and Postlewaite(2017)]{mclean2017dynamic}
Richard~P McLean and Andrew Postlewaite.
\newblock A dynamic non-direct implementation mechanism for interdependent
  value problems.
\newblock \emph{Games and Economic Behavior}, 101:\penalty0 34--48, 2017.

\bibitem[Narahari(2014)]{narahari2014game}
Yadati Narahari.
\newblock \emph{Game theory and mechanism design}, volume~4.
\newblock World Scientific, 2014.

\bibitem[Perry and Reny(2002)]{perry2002efficient}
Motty Perry and Philip~J Reny.
\newblock An efficient auction.
\newblock \emph{Econometrica}, 70\penalty0 (3):\penalty0 1199--1212, 2002.

\bibitem[Roth(1989)]{roth1989two}
Alvin~E Roth.
\newblock Two-sided matching with incomplete information about others'
  preferences.
\newblock \emph{Games and Economic Behavior}, 1\penalty0 (2):\penalty0
  191--209, 1989.

\end{thebibliography}

\end{document}